\documentstyle[aps,prd,amssymb]{revtex}
\input epsfig.sty

\draft

\begin{document}

\title{ Magnetic knots as the origin of spikes in the gravitational
wave backgrounds}

\author{ Massimo Giovannini\footnote{
electronic address: M.Giovannini@damtp.cam.ac.uk} }

\address{{\it DAMTP,  Silver Street, CB3 9EW Cambridge,
United Kingdom}}

\maketitle
\begin{abstract}
The dynamical symmetries of  hot and electrically neutral plasmas in a
highly conducting medium suggest that, after the epoch of the 
electron-positron annihilation, 
 magnetohydrodynamical configurations carrying a net magnetic
helicity can be present. The simultaneous conservation of the
magnetic flux and helicity implies
that the (divergence free) field lines will possess {\em inhomogeneous}
 knot structures  acting  as source {\em ``seeds''} 
in the evolution equations of the scalar,
vector and tensor fluctuations of the background geometry. We give
explicit examples of magnetic knot configurations with finite energy
and we compute the induced metric fluctuations. 
Since magnetic knots are (conformally)
 coupled to gravity via the vertex
dictated by the equivalence principle, they
can imprint spikes in the gravitational wave spectrum for frequencies 
compatible with the typical scale of the
knot corresponding, in our examples, to a present frequency  range of
$10^{-11}$--$10^{-12}$ Hertz. At lower frequencies the spectrum is
power-suppressed and well below the COBE limit. For smaller length
scales (i.e. for larger frequencies)
 the spectrum is exponentially suppressed  and then irrelevant
for the pulsar bounds. Depending upon the number of knots of the
configuration, the typical amplitude of the gravitational wave
logarithmic energy spectrum (in critical units)
can be even four orders of magnitude larger than the usual
flat (inflationary) energy spectrum generated thanks to the
parametric amplification of the vacuum fluctuations.
\end{abstract}
\vskip0.5pc
\centerline{Preprint Number: DAMTP-1998-107\\
{\em to appear in Physical Review D}}
\vskip0.5pc
\noindent

\renewcommand{\theequation}{1.\arabic{equation}}
\setcounter{equation}{0}
\section{Introduction}
The idea that large scale magnetic fields could exist in the early
Universe is both appealing and plausible \cite{Z1}.
It was noticed long
ago that if magnetic fields existed in the early Universe, they could
have affected (or possibly explained)  various stages of
the formation of cosmic structures \cite{R,Z2}.
Long range stochastic fields can modify the rate of the Universe
expansion  and the reaction rate at  the nucleosynthesis epoch 
\cite{ker}. Stochastic fields coherent over the horizon size at the
decoupling time can depolarise the Cosmic Microwave Background Radiation
(CMBR) \cite{pol},
and can also change the sound velocity of the plasma at the
corresponding era \cite{gr2}.

If the temperature of the radiation-dominated Universe
is higher than the temperature of the
electroweak phase transition (i.e. $T> 100 ~~{\rm GeV}$), then the
large scale  components of the standard model gauge fields 
should be identified with the hypermagnetic fields \cite{mis} 
which, by contributing to
the (Abelian) anomaly, 
can generate matter--antimatter fluctuations \cite{mis1} (possibly
relevant in the context of the inhomogeneous nucleosynthesis scenario 
\cite{IBBN,revN}). Long range hypermagnetic fields can also
influence the dynamics of the phase transition itself \cite{mis3}.
Interesting consequences of the presence of 
electroweak condensates for the gravitational wave 
spectra  were also discussed in \cite{rub}.

Stochastically distributed
magnetic fields (either amplified from their vacuum fluctuations
\cite{mgg}
through the breaking of the conformal invariance of Maxwell equations
or just assumed \cite{Ba1}) can  imprint inhomogeneities in the
microwave
sky. Long ago it was argued that any uniform (spatially
homogeneous) magnetic fields can make the energy momentum tensor of a
radiation dominated universe slightly anisotropic producing,
ultimately,  an
observable effect in the microwave sky  \cite{Z2}. Recently bounds
were derived (using the COBE 4-yrs maps)  on  the present strength of
a uniform (and homogeneous) magnetic field \cite{Ba2}.

In this paper we will argue that,  on the basis of the physical laws
governing the evolution of the mean magnetic fields in a hot plasma,
there are no reasons why the topology of the magnetic flux lines
should be {\em trivial}, namely there are no reasons why the magnetic
flux lines should not intersect each others. If the topology of the
magnetic flux lines is non trivial, then, the helicity of a specific
magnetic configuration will also enter in the  equations
describing evolution of the scalar,
vector and tensor fluctuations of the metric in a radiation dominated
Universe. Provided the typical scale of the knot (red-shifted at the
recombination time) is much smaller that the magnetic Jeans scale at
the corresponding epoch, very little impact can be expected for
structure formation \cite{was,pee}.

In short our logic is the following. One of the main features of
the evolution equations of the mean
magnetic fields (for scales larger than the Debye radius)
  in  highly conducting (and globally neutral) plasmas
is that they are divergence-less (i.e.  the field lines have no
end points) and, therefore, the structures arising in
magnetohydrodynamics (MHD) can be examined in terms of the topology of
closed curves \cite{ber,bis}. Now, it is well known that in (ideal)
MHD a unique and intriguing property holds: the {\em magnetic flux}  and
{\em magnetic helicity} are simultaneously conserved all along the
evolution of the plasma for scales larger than the magnetic
diffusivity scale. Since the Universe is a good conductor
all along its history, we have to conclude that, on one hand,
 thanks to the magnetic flux conservation,  the
magnetic flux lines always evolve glued together with the plasma
element and, on the other
hand, thanks to the magnetic helicity conservation, the sum
of  the number of {\em twists} and the number of {\em knots}
 in the flux lines is also left  invariant \cite{ber} as time goes by.
At this point the choice is twofold. Either we assume (as it is
frequently   done
in the case of stochastic magnetic backgrounds \cite{mgg,seed}) 
that the (mean)
magnetic helicity is strictly zero, or we accept that the magnetic
helicity can be present and, in this last case, 
it will be conserved to a very good approximation. 

As we said, a non-vanishing magnetic helicity implies that the
corresponding magnetic field  
configurations must be inhomogeneous
 (otherwise the integral defining the magnetic helicity 
cannot be gauge-invariant). Thus, in the present paper, we are forced
to  deal with inhomogeneous fields. At the same time one could
say that inhomogeneous configurations carrying zero (net) magnetic
helicity can also have interesting effects on the gravitational wave
production. This is certainly true and one of the purposes of the
present investigation is to understand, with some specific examples,
what difference does it make to drop the (oversimplifying) assumption
that the magnetic flux lines of  the cosmological magnetic fields have
all trivial topological structure.

Inhomogeneous magnetic knots are  expected to affect various aspects of the
life of our Universe. They can represent a quite
interesting source ``seed'' in the evolution  of metric density
perturbations during  radiation (and possibly matter) dominated
epochs.  Since  magnetic
knot configurations with finite energy and helicity are necessarily
inhomogeneous, no homogeneous and isotropic (background) contribution
is expected from the knot energy-momentum tensor.
At the same time, the presence of  magnetic knots and twists in the
magnetic flux lines does modify the evolution equations of the 
 metric fluctuations.
The spatial components  of the knot energy-momentum tensor form
a three-dimensional rank two (cartesian) tensor which can be
 decomposed into irreducible  scalar
(invariant under three dimensional rotations), vector and  tensor
parts providing source terms for the evolution equations of the
corresponding scalar vector and tensor parts of the metric
inhomogeneities. The tensor modes of the metric are automatically
invariant under infinitesimal coordinate (gauge) transformations
\cite{bard}. The components of the knot energy-momentum
tensor are also invariant under infinitesimal coordinate
transformations because of the absence of any (homogeneous and isotropic)
magnetic background.

We want to notice immediately that, one of the main
ideas of our investigation is the connection among knot seeds and metric
fluctuations. Up to now all the investigations on the existence of
large scale magnetic fields were mainly devoted to the study of
suitable mechanisms able to explain the large scale ``magnetic"
structure of the Universe, or more specifically, the existence of
galactic (and possibly inter-galactic) magnetic fields \cite{seed}.
In our study we point out that, if the conservation properties
arising in MHD are properly taken into account, then, interesting
effects can be foreseen also for the metric fluctuations.

In the first place we would like to motivate magnetic knots
configurations on the basis of magnetohydrodynamics (MHD).
The idea of magnetic knots might seem, at first sight, a crazy
one. In order to make this idea quantitatively plausible we will
introduce some explicit MHD solutions describing magnetic knots. It is
very important that these field configurations are regular over the
whole space, they have finite energy and helicity. They represent
then a good framework where the effect of magnetic knots on the
metric inhomogeneities can be discussed.
The fact that our configurations have finite energy (as also required
by considerations related to the $U_{em}(1)$ gauge invariance of the
magnetic helicity) implies that the amplitude of the magnetic fields
tends to zero for scales larger than the scale of the configuration
$L_{s}$. For the discussion of  the  magnetic helicity evolution
 in a  radiation dominated stage of expansion an important tool
will be the conformal invariance of the (ideal) MHD equations.
Moreover,  since we want our MHD description to be fully valid we
will also consider times larger than the epoch of electron-positron
annihilation.

The second purpose of this investigation are the possible
phenomenological consequences of the existence of magnetic knots
prior to the decoupling time.
We will show that spikes in the gravitational waves energy spectrum 
\cite{revGW}
can be foreseen and  we will compare our results with the typical
amplitudes of gravitational waves produced in the context of
inflationary models.
The amplitude of the ``magnetic" spikes depends directly upon the
helicity of our configuration  and turns out to be substantially
larger than the signal of a stochastic gravitational wave background
with flat (Harrison-Zeldovich) logarithmic energy spectrum which
represents a generic prediction of ordinary inflationary models.
The ``magnetic spikes" in the gravitational wave spectrum are not
constrained by pulsars's pulses since they are expected to occur in a
(present) frequency range $\omega_{dec}<\omega<\omega_{e^{+}e^{-}}$
(where $\omega_{dec} \sim 10^{-16}~{\rm Hz}$ and $\omega_{e^{+}
e^{-}}\sim 10^{-10}~{\rm Hz}$.
In principle we could also expect interesting effects from the
Lorentz force term associated with the knot configuration and
appearing in the Euler (Navier-Stokes) equation for the density field
\cite{was,pee} which goes, in MHD, as $\vec{J}
\times \vec{H} \sim (\vec{\nabla}\times\vec{H})\times\vec{H}$. The
calculation shows that, since the knot scale is just slightly smaller
than the magnetic Jeans scale at the decoupling , effects on
structure formation are expected to be mild.

The plan of our paper is then the following. In Sec. II we will
recall
the basic equations and definitions with particular attention to the
Alfv\'en theorem and to the magnetic helicity theorem. In Sec. III we
will introduce the class of configurations with finite energy and
helicity.
In Sec. IV we will introduce the gravitational effects of the knots
and in Sec. V we will focus our attention on the problem of magnetic
spikes in the gravitational wave spectrum. Section VI contains our
concluding remarks. We made the choice of including in the Appendix
different technical results which might be of help for the interested
reader.

\renewcommand{\theequation}{2.\arabic{equation}}
\setcounter{equation}{0}
\section{Dynamical symmetries in hot plasmas }

In a conformally flat metric of Friedman-Robertson-Walker (FRW) type
the line element is
\begin{equation}
ds^2 = a^2(\eta) \bigl[ d\eta^2 - d\vec{x}^2\bigr],~~~a(\eta)
\sim\eta.
\end{equation}
Since we are going to exploit the Weyl invariance of the Maxwell
 fields in this type of  backgrounds we stress, for our future
convenience,
 that $(\eta,\vec{x})$ are the conformal coordinates which are
related
 to the  flat coordinates $(t,\vec{y})$ by the 
 relations $d\eta= dt/a(t),~~~d{\vec{x}}= d\vec{y}/a(t)$. In the
 following we will denote by a prime the derivation with respect to
 conformal time and, when needed (for instance in Sec. III), by the
 over-dot the derivation with respect to the cosmic time $t$.

Using the magnetohydrodynamical approximation the
evolution equations of Maxwell fields at finite conductivity and
 diffusivity have to be supplemented by the Navier-Stokes
equation. By exploiting conformal invariance they can be written as
\cite{mis1,bis,mhd}
\begin{eqnarray}
&&\biggl[\biggl(p + \rho\biggr)\vec{v}\biggr]' +
\vec{v}\cdot\vec{\nabla}\biggl[\biggl(p+ \rho\biggr)\vec{v}\biggr] +
\vec{v}~\vec{\nabla}\cdot\biggl[\biggl(p+ \rho\biggr)\vec{v}\biggr]
=-\vec{\nabla} p  + \vec{J}\times\vec{H} + \nu \biggl[\nabla^2
\vec{v}
+ \frac{1}{3} \vec{\nabla}(\vec{\nabla}\cdot\vec{v})\biggr],
\label{ns}\\
&&\vec{H}' + \vec{\nabla}\times\vec{E} =0,~~~\vec{\nabla}
\cdot\vec{E}=0,
\label{Mx1}\\
&&\vec{\nabla}\times\vec{H}= \vec{J} + \vec{E}',~~~\vec{\nabla}
\cdot\vec{H}=0,
\label{Mx2}\\
&&\vec{J} = \sigma \biggl(\vec{E} + \vec{v}\times \vec{H}\biggr),
\label{ohm}
\end{eqnarray}
where $\vec{v} = \vec{x}~'(\eta)$ is the bulk velocity of the plasma,
$\sigma$ is the conductivity and $\nu$ is the shear
viscosity coefficient;
the curved space quantities appearing in Eqs. (\ref{ns})-(\ref{ohm})
are related to the flat space ones by the  metric rescaling
\begin{equation}
\vec{E} = a^2 \vec{{\cal E}},~~\vec{H} = a^2 \vec{{\cal
B}},~~\sigma=a
\sigma_{c},~~ \rho =
a^4 \rho_{c},~~p= a^4 p_{c},~~\vec{J} = a^3 \vec{j},~~\nu =
a^3\nu_{c}.
\label{defin}
\end{equation}
Notice that in Eqs. (\ref{Mx1}) $\vec{\nabla}\cdot\vec{E} =0$ is not
imposed as an additional equation on the fluid and indeed is not a
valid equation for compressional waves which emerge from the theory
\cite{bis,kat}. However, if the plasma is globally neutral on scales
much larger than the Debye radius (i.e. $L_{D}(T)\sim \sqrt{T/(n_{e}
e^2)}$ where $n_{e}$ is the average electron density )
 the mean electric fields are effectively divergence free. After  the
epoch of $e^{+}$--$e^{-}$ annihilation (i.e. $T <0.1$ MeV) we have
that $n_{e} \sim 6.43 \times 10^{-9}~ x_e~ \Omega_{B}h_{100}^2 ~T^3$,
 and it turns out that
$L_{D}(T) \sim 1.9 \times 10^{-4} (0.1 ~{\rm MeV}/T) ~{\rm cm}$
($x_{e}$ is the ionization fraction and $\Omega_{B}$ the baryon
density in critical units). If
we compare the Debye scale with the horizon distance at $T\sim
0.1~{\rm MeV}$ we have that $L_{H}(t_{e^{+}e^{-}}) \sim 10^{12}~{\rm
cm}\gg L_{D}(t_{e^{+}e^{-}})$. Therefore, for our purposes the
plasma description is indeed appropriate.

The set of MHD equations can be studied with different
``closures''. Some usual closures \cite{bis}
are  the incompressibility (i.e. $\vec{\nabla}\cdot\vec{v}=0$) and
the
adiabaticity, but other closures can be invented depending upon the
problem at hand
(like the isothermal closure or the closure $\vec{J}
={\rm constant}$). Since the evolution of the Universe can be
described
(within the hot Big-Bang model) by an adiabatic expansion, we can
certainly require that $\rho =a^4 \rho_{c} \sim {\rm constant}$ and
that $p=
\rho/3$. Moreover one can also assume that the fluid is
incompressible and in this case, not only the magnetic lines will be
divergence-less but also the velocity lines will share the same
property. We notice, incidentally, that in the incompressible
approximation we can also define a hydrodynamical helicity \cite{mof}
which allows the topological treatment also for the hydrodynamical
part of the system. At the end of Sec. IV we will specifically
consider a case where the incompressible closure is relaxed.

In referring the reader to Appendix A for further details concerning
 the MHD evolution we want only to note that at high conductivity 
the Alfv\'en theorem
holds
\begin{eqnarray}
\frac{d}{d\eta} \int_{\Sigma} \vec{H} \cdot d\vec{\Sigma}=-
\frac{1}{\sigma} \int_{\Sigma} \vec{\nabla} \times\vec{\nabla}
\times\vec{H}\cdot d\vec{\Sigma},
\label{flux}
\end{eqnarray}
where $\Sigma$ is an arbitrary closed surface which moves together with the
plasma.
If we are in the inertial regime (i.e. $L>L_{\sigma}$ where
$L_{\sigma}$ is the magnetic diffusivity length), we can
say that the expression appearing at the right hand side is
subleading
and the magnetic flux lines evolve glued to the plasma element.
Moreover,  the magnetic helicity  ${\cal H}_{M}$ is conserved in the
inertial range:
\begin{equation}
{\cal H}_{M} = \int_{V} d^3 x \vec{A}~\cdot \vec{H},
~~~\frac{d}{d\eta} {\cal H}_{M} = - \frac{1}{\sigma} \int_{V} d^3 x
{}~\vec{H}\cdot\vec{\nabla} \times\vec{H}.
\label{h2}
\end{equation}
Concerning Eqs. (\ref{h2}) two comments are in
order. In Eq. (\ref{h2}) the vector potential appears and, therefore
it might seem that the expression is not gauge invariant. This is not
the case. In fact $\vec{A}\cdot\vec{H}$ is not gauge invariant but,
nonetheless, ${\cal H}_{M}$ {\em is gauge-invariant} since we will define
the integration volume in such a way that the magnetic field
$\vec{H}$
is parallel to the surface which bounds $V$ and which we will call
$\partial V$. Calling $\vec{n}$ is the unit vector normal to $\partial
V$, the integral defining the magnetic helicity is gauge invariant
provided $\vec{H}\cdot\vec{n}=0$ in $\partial V$. 
As we will consider finite
energy fields in our examples of Section III we will have that
$\vec{H}=0$ in any part of $\partial V$ 
(which might be a closed surface at infinity).
In Eq. (\ref{h2}) the term appearing under integration at the right
hand side (i.e. $\vec{H} \cdot \vec{\nabla}\times {\vec{H}}$) is
sometimes called also magnetic helicity (or magnetic gyrotropy) and it
is a gauge invariant measure of the diffusion rate of ${\cal H}_{M}$
at finite conductivity. The two theorems given in
Eqs. (\ref{flux})--(\ref{h2}) are reviewed and  proven in Appendix A.

Notice that Eqs. (\ref{flux}) and (\ref{h2}) are truly dynamical symmetries of
the
plasma, in the sense that they are not inherent to static MHD
configurations.
Before ending this section we want to mention a further scale which
turns out to be important in the discussion of the large scale
effects of primordial magnetic fields, namely the magnetic Jeans
scale \cite{was,pee}.
At $t= t_{dec}$
we have from Eq. (\ref{diffscale}) that $L_{\sigma}(t_{dec}) \simeq
4.7 ~\times 10^{10}~{\rm cm}$ whereas the magnetic Jeans scale is, at
the same epoch,
\begin{equation}
L_{B_{J}} \simeq \frac{|\vec{{\cal
B}}(t_{dec})|~M_{P}}{\rho_{c}(t_{dec})}.
\label{jeans}
\end{equation}
In a matter-dominated Universe the magnetic Jeans scale
evolves like the scale factor.
In order to make our estimate more constrained we can assume
 a field sufficiently strong in order to rotate the
polarisation plane of the CMBR at the decoupling \cite{pol}
(i.e. $|\vec{H}(t_{dec})|\geq 10^{-3}$ ) and we see that
\begin{equation}
L_{B_{J}}(t_{dec}) \simeq 10^{21} ~{\rm cm}.
\end{equation}
{}From this last numerical estimate we see that if the magnetic field
is sufficiently strong, the magnetic Jeans length is of the same
order of the horizon size  (i.e. $L_{H}(t_{dec}) \sim
L_{B_{J}}(t_{dec})) $. This result has interesting implications for
the possible impact of magnetic field on structure formation. In
particular, we have that if $L_{B_{J}}(t_{dec}) \sim L_{H}(t_{dec})$,
then the density contrast induced by the Lorentz force term of Eqs.
(\ref{ns}) and (\ref{ns3}) can be of order 1 at the time of galaxy
formation \cite{was}. In the opposite case (i.e. $L_{B_{J}}(t_{dec})
\ll L_{H}(t_{dec})$), from the Lorentz force term we cannot expect
any significant effect on structure formation. In Sec. V we will come
back to this problem.

\renewcommand{\theequation}{3.\arabic{equation}}
\setcounter{equation}{0}
\section{Magnetic knot configurations}

Since both magnetic helicity and magnetic flux are conserved during
the dynamical evolution in the inertial range, 
configurations with non trivial topological structure present at 
some initial time  will be preserved all along the dynamical evolution.

In this Section we will provide some examples of magnetic knot
configurations. In order to perform various calculations, it is useful
to employ configurations which are non singular at large and small
distances. In this way all the integrals defining the helicity, the
gyrotropy and the total energy. will be automatically well defined.
This is exactly what happens in the following examples.
Consider the  magnetic field with (spherical) components
\begin{equation}
H_{r}({\cal R},\theta,n) = - \frac{4 B_{0}}{\pi~ L_{s}^2}\frac{n
\cos{\theta}}{\bigl[  {\cal
R} + 1\bigr]^2},~~~
H_{\theta}({\cal R}, \theta,n) = - \frac{4 B_{0}}{\pi~
L_{s}^2}\frac{{\cal R}^2 -1}{\bigl[
{\cal R}^2 + 1\bigr]^3}n \sin{\theta},~~~
H_{\phi}({\cal R}, \theta) = - \frac{8 B_0}{ \pi~ L_{s}^2}\frac{ r
\sin{\theta}}{\bigl[
{\cal R}^2 + 1\bigr]^3}
\label{knot}
\end{equation}
\begin{figure}
\centerline{\epsfxsize = 12 cm  \epsffile{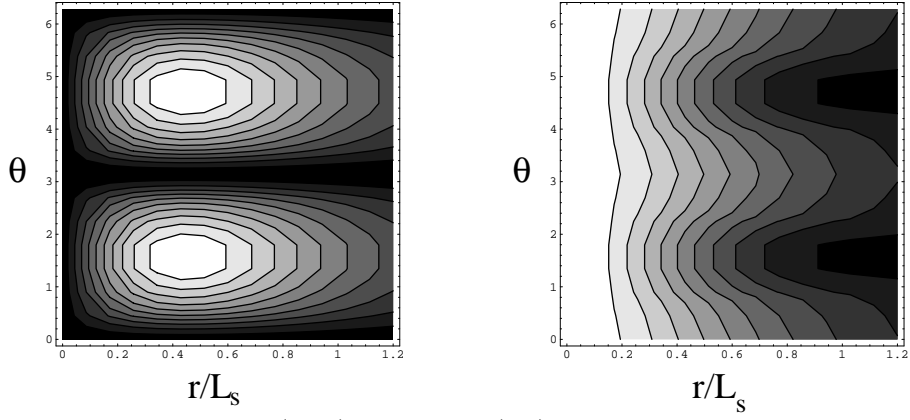}}
\caption[a]{We plot the moduli of the poloidal (right) and toroidal
(left) components of the field configurations reported in
Eq. (\ref{knot}). In these topographic maps dark regions correspond
to low intensity spots of the field whereas brighter regions
correspond to higher intensities.
The toroidal component does not depend on $n$.
The poloidal component is instead computed in the case of $n=
100$. Moreover in both case we took $|\vec{H}_0(t_{dec})| \sim 0.001$
Gauss. Notice that in the limit of $n\rightarrow 0$ the poloidal
component goes to zero. For $r = L/L_{s} >2$ the field intensity is
suppressed.}
\label{cont}
\end{figure}
In Eq. (\ref{knot}) $L_s$ represent the typical scale of the seed and
${\cal R}= r/L_{s}$. Moreover $B_{0}= H_0 L_{s}^{2}$  is the
dimensionless
amplitude of the magnetic field; $n$ is just an integer number.
It can be also intuitively useful to report explicitly the moduli of
the poloidal and toroidal components of $\vec{H}$
\begin{equation}
|\vec{H}_{p}({\cal R}, \theta,n)| = \frac{4 B_0} { \pi~ L_{s}^2} n
\frac{ \sqrt{ {\cal R}^4 + 2
{\cal R}^2 \cos{2 \theta}  +1 }}{ \bigl[ {\cal R}^2 + 1 \bigr]^3},~~~~~
|\vec{H}_{t}({\cal R}, {\theta})| = \frac{ 8 B_0} { \pi~
L_{s}^2}\frac{ {\cal R} \sin{\theta}
}{  \bigl[ {\cal R}^2 + 1 \bigr]^3}
\label{poloi}
\end{equation}
where the poloidal and toroidal components of the field are defined
in
the standard way ($ \vec{H}_{p} = H_{r} \vec{e}_{r} + H_{\theta}
\vec{e}_{\theta}$, $\vec{H}_{t}= H_{\phi} \vec{e}_{\phi}$). The
configurations (\ref{knot}) were firstly introduced in \cite{ran} in
the context of the study of the topological properties of the
electromagnetic flux lines.
\begin{figure}
\centerline{\epsfxsize = 12 cm  \epsffile{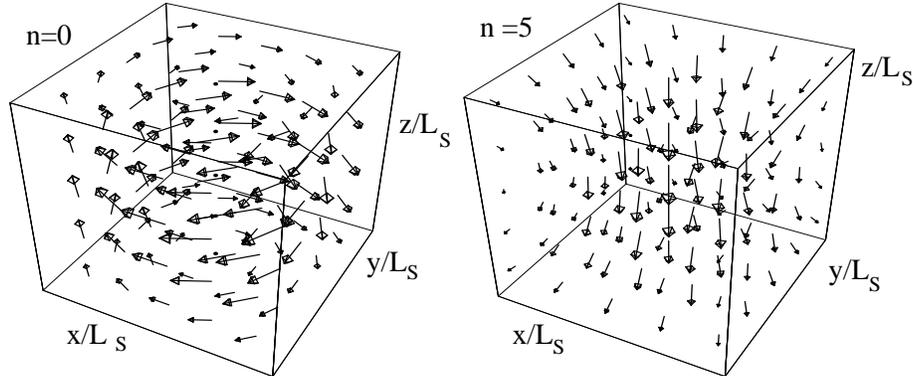}}
\caption[a]{ We plot the magnetic field in Cartesian components (see
Eqs. (\ref{cart}) in Appendix B) for the cases $n=0$ (zero helicity)
and $n= 5$. The
direction of the arrow in each point represents the tangent to the
flux lines of the magnetic field, whereas its length is proportional
to
the field intensity. We notice that for large $n$ the field is
concentrated, in practice,  in the core of the
knot (i.e. $r/L_{s} <1$). This can be also argued
from Fig. (\ref{cont}).}
\label{FIG1}
\end{figure}
{}From the  explicit expressions of the vector potentials
\begin{eqnarray}
A_{r}({\cal R},\theta) &=& - \frac{2 B_0}{ \pi L_{s}}
\frac{\cos{\theta} }{\bigl[{\cal R}^2 +1\bigr]^2},~~~ {\cal
A}_{r}({\cal R}, \theta,\eta)= \frac{A_{r}({\cal
R},\theta)}{a(\eta)},
\nonumber\\
A_{\theta}({\cal R},\theta) &=& \frac{2 B_0}
{ \pi L_{s}} \frac{ \sin{\theta}}{\bigl[ {\cal
R}^2 + 1\bigr]^2},~~~{\cal A}_{\theta}({\cal R}, \theta,
\eta)=\frac{A_{\theta}({\cal R}, \theta)}{a(\eta)},
\nonumber\\
A_{\phi}({\cal R},\theta,n) &=& - \frac{ 2 B_0}{ \pi L_{s}} \frac{ n
{\cal
R}\sin{\theta}}{\bigl[{\cal R}^2 + 1\bigr]^2},~~~ {\cal
A}_{\phi}({\cal R},\theta,\eta) = \frac{A_{\phi}({\cal
R},\theta)}{a(\eta)},
\end{eqnarray}
the magnetic helicity can be computed
\begin{equation}
{\cal H}_{M} =\int_{V} \vec{A} \cdot \vec{H} d^3 x= \int_{0}^{\infty}
\frac{ 8 n
B^2_0}{\pi^2} \frac{ {\cal R}^2 d {\cal R}}{\bigr[ {\cal R}^2 +
1\bigl]^4} = n B^2_0.
\end{equation}
Notice that the integration is perfectly convergent over the whole
space and, since the field is $\vec{H}=0$ in any part of $\partial V$
at infinity, this quantity is also gauge invariant.
We can compute also the total energy of the field, namely
\begin{equation}
E = \frac{1}{2}\int_{V} d^3 x |\vec{H}|^2 = \frac{B^2_0}{2~L_{s}}
(n^2 + 1).
\end{equation}
The magnetic configurations we just presented are solutions of the
MHD equations (in the inertial range) with large Prandtl number and
small $\beta$ parameter ($\beta \ll 1,~~~Pr_{M} \gg 1$) for a velocity
field $\vec{v} =\pm {\vec{H}}/{\bigl[\rho + p\bigr]}$ (see Appendix A
for more detailed explanation of these notations).

In the limit $n\rightarrow 0$ the total helicity goes to zero and, as
a consequence, the magnetic field is purely toroidal.
In the case where the helicity is non vanishing we can also compute
the volume integral of the magnetic gyrotropy $\vec{H} \cdot
\vec{\nabla} \times\vec{H}$ which, according to Eq. (\ref{h2}),
measures the diffusion rate of ${\cal H }_{M}$ at finite
conductivity. From Eq. (\ref{knot}) we have that
\begin{equation}
\int_{V} \vec{H} \cdot \vec{\nabla} \times \vec{H} d^3 x=
\frac{256~B^2_0~n}{\pi L^2 } \int_{0}^{\infty} \frac{ {\cal R}^2 d
{\cal R}}{(1 + {\cal R}^2)^5} = \frac{5 {\cal H}_{M}}{L_s^2}.
\end{equation}
Using this last result into Eq. (\ref{h2}) we obtain the explicit
version of the helicity dilution equation valid in the context of our
configurations, namely
\begin{equation}
\frac{d}{d\eta} {\cal H}_{M} = - \frac{5}{\sigma ~L_{s}^2} {\cal
H}_{M}.
\label{dil}
\end{equation}

\renewcommand{\theequation}{4.\arabic{equation}}
\setcounter{equation}{0}
\section{Gravitational waves and magnetic knots}

Magnetic knot configurations present right after the epoch of
electron-positron annihilation can act as seed sources for the
evolution equations of the scalar, vector and tensor metric
fluctuations. In fact, the spatial part of the knot energy momentum
tensor (computed in Eqs. (\ref{enmom}) of Appendix B) is a symmetric
three-dimensional tensor of rank two with six degrees of freedom
which
transform differently under three-dimensional rotations on the $\eta=
constant$ hypersurface. More precisely the spatial components of the
knot energy-momentum tensor can be decomposed in a trace part plus a
traceless scalar with one (scalar) degree of freedom each, and vector
and tensor parts with two degrees of freedom each. The  energy
momentum tensor of the knot has no homogeneous background
contribution. This fact has two important consequences. On one hand
the knot does not affect the background evolution of a radiation (or
matter) dominated Universe, on the other hand  the energy momentum
tensor is automatically invariant for infinitesimal (gauge)
coordinate transformations. In the following discussion we will focus
our attention on the evolution of the tensor modes. The tensor modes
of the metric correspond to gravitational waves \cite{revGW}.

Consider then a perturbation of a (homogeneous and isotropic)
conformally flat metric
\begin{equation}
g_{\mu\nu} \rightarrow g_{\mu\nu}(\eta) + \delta
g_{\mu\nu}(\vec{x},\eta).
\end{equation}
whose fluctuating part, $\delta g_{\mu\nu}$, contains scalar, vector 
and tensor modes
\begin{equation}
 \delta g_{\mu\nu}(\vec{x},\eta) =   \delta
 g^{(S)}_{\mu\nu}(\vec{x},\eta)+  \delta
 g^{(V)}_{\mu\nu}(\vec{x},\eta) +  \delta
 g^{(S)}_{\mu\nu}(\vec{x},\eta),
\end{equation}
The corresponding equations of motion will have, as source term,
 the scalar vector and
tensor modes of the knot energy-momentum tensor
\begin{equation}
\delta T_{\mu\nu}(\vec{x},\eta) = \delta
T^{(S)}_{\mu\nu}(\vec{x},\eta) +
\delta T^{(V)}_{\mu\nu}(\vec{x},\eta) +  \delta
T^{(T)}_{\mu\nu}(\vec{x},\eta) .
\end{equation}
Tensor perturbations of the metric are constructed using a symmetric
three-tensor which satisfies the constraints
\begin{equation}
h_{i}^{i} = \nabla_{i} h^{i}_{j} =0,
\label{T1}
\end{equation}
which implies that $h_{ij}$ does not contain parts transforming as
scalars or vectors. Therefore the perturbed tensor components of the
metric will be
\begin{equation}
\delta g^{T}_{00}=0,~~\delta^{(T)}_{\mu 0}=0,~~ \delta g_{ij} = -
a^2(\eta) h_{ij}(\vec{x},\eta),
\label{T2}
\end{equation}
and the perturbed line element becomes
\begin{equation}
ds^2 = a^2(\eta)\biggl[ d\eta^2 - (\gamma_{ij} + h_{ij}) d x^{i}d
x^{j} \biggr],
\label{einp}
\end{equation}
where $\gamma_{ij}$ denotes the spatial part of the background metric.
The evolution equations for the tensor  modes of the geometry
 can then be written as
\begin{equation}
\delta G^{(T)}_{\mu\nu} = 8\pi G \delta T^{(T)}_{\mu\nu},
\end{equation}
where $\delta G^{(T)}_{\mu\nu}$ are the tensor components of the 
Einstein tensor $G_{\mu \nu} = R_{\mu\nu} - \frac{1}{2} g_{\mu\nu} R$
perturbed according to Eqs. (\ref{T1})--(\ref{T2}).
Eq. (\ref{einp}) equation becomes, in conformal time,
\begin{equation}
h_{ij}'' + 2 {\cal H} h_{ij}' - \nabla^2 h_{ij} = - 16 \pi G
\tau^{(T)}_{ij}.
\label{GWeq}
\end{equation}
Notice that the energy-momentum tensor can be decomposed, in Fourier
space, as
\begin{eqnarray}
&&\tau_{ij}(\vec{x}) = \frac{1}{(2\pi)^{\frac{3}{2}} }\int
e^{i\vec{k}\cdot\vec{x}} \tau_{ij}(\vec{k}) d^3 k,~~~\tau(\vec{k})
\equiv \tau_{i i} (\vec{k}),
\nonumber\\
&&\tau_{ij}(\vec{k})= \frac{1}{3} \delta_{ij} \tau(\vec{k}) + \biggl(
\hat{k}_{i}
\hat{k}_{j} - \frac{1}{3} \delta_{ij} \tau^{(S)}(\vec{k})\biggr) +
\biggl( \hat{k}_{i}
\tau^{(V)}_{j}(\vec{k}) + \hat{k}_{j} \tau^{(V)}_{i}(\vec{k})\biggr)
+
\tau^{(T)}_{ij}(\vec{k}),~~~\hat{k}_{i} = \frac{k_{i}}{|\vec{k}|},
\end{eqnarray}
where
\begin{eqnarray}
&&\tau^{(S)}(\vec{k}) = \frac{3}{2} \hat{k}_{i} \hat{k}_{j}
\tau_{ij}(\vec{k}) - \frac{1}{2}\tau(\vec{k}),
\nonumber\\
&&\tau^{(V)}_{i}(\vec{k}) = \tau_{i m}(\vec{k}) \hat{k}_{m}
- \hat{k}_{i} \hat{k}_{j}\tau_{j m}(\vec{k}) \hat{k}_{m},
\nonumber\\
&&\tau^{(T)}_{ij}(\vec{k}) = \tau_{ij}(\vec{k}) + \frac{1}{2} \biggl(
\hat{k}_{i}\hat{k}_{j} - \delta_{ij}\biggr)\tau(\vec{k}) +
\frac{1}{2}
\biggl(\hat{k}_{i} \hat{k}_{j} + \delta_{ij} \biggr) \hat{k}_{m}
\hat{k}_{n} \tau_{m n}(\vec{k}) - \hat{k}_{i} \hat{k}_{m} \tau_{m j}(\vec{k})
-
\hat{k}_{m} \hat{k}_{j} \tau_{i m}(\vec{k}),
\label{decompposition}
\end{eqnarray}
In Appendix C the Fourier transforms of each component of the energy
momentum tensor is reported. Here we will simply discuss the results
of
the expansions in the two  limits relevant for
physical applications. For frequencies much larger than the typical
frequency of the knot (i.e. $k> k_{s} \sim L_{s}^{-1}$)  we have that
the
Fourier transform of each component of the energy momentum tensor is
exponentially suppressed  $ \exp{[-k/k_{s}]}$, whereas for small
frequencies (infra-red limit) the various components have different
power-law behaviours:
\begin{eqnarray}
&&\tau_{xx}(\vec{k}) \sim \tau_{yy}(\vec{k}) = \frac{ B_0^2}{\pi^2
L_{s}^4 k^3_{s}} \sqrt{\frac{\pi}{2}}\frac{n^2}{5} \biggl[ 1 + \kappa
+ {\cal
O}\biggl(\biggl(\frac{k}{k_s}\biggr)^2\biggr) \biggr] e^{-
\frac{k}{k_{s}}},
\nonumber\\
&&\tau_{zz}(\vec{k}) = \frac{ B_0^2 }{\pi^2 k^3_{s}
L_{s}^4} \sqrt{\frac{\pi}{2}}\frac{5-3 n^2}{20} \biggl[ 1 + \kappa +
{\cal O}\biggl(\biggl(\frac{k}{k_s}\biggr)^2\biggr) \biggr]
e^{-\frac{k}{k_s}},
\nonumber\\
&&\tau_{xz}(\vec{k}) = \frac{ B_0^2 }{\pi^2 k^3_{s}L_{s}^4} \biggl[{\cal
O}\biggl(\frac{k}{k_s}\biggr)\biggr]
 e^{-\frac{k}{k_s}},~~~\tau_{y z}(\vec{k}) = \frac{ B_0^2 }{\pi^2 k^3_{s}
L_{s}^4} \biggl[{\cal O}\biggl(\frac{k}{k_s}\biggr)\biggr] e^{-\frac{k}{k_s}} ,
{}~~~\tau_{x y}(\vec{k}) = \frac{ B_0^2 }{\pi^2 k^3_{s}L_{s}^4} \biggl[{\cal
O}\biggl(\biggl(\frac{k}{k_s}\biggr)^2\biggr)\biggr] e^{-\frac{k}{k_s}},
\label{expan}
\end{eqnarray}
( ${\cal O}$ denotes  the Landau symbol; notice that, within the
square
brackets we gave the infra-red expansion of the Fourier transform,
whereas, outside the brackets we kept the leading [ultra-violet] exponential
suppression which always factorizes according to the exact results
reported in Appendix C). From Eqs. (\ref{expan}) we can see
 that the off-diagonal terms are
always subleading (at large scales) if compared with the diagonal
ones.

{}From Eq. (\ref{GWeq}) it  is possible to define the (Fourier space)
evolution of the two transverse and traceless degrees (metric) of
freedom which reads
\begin{eqnarray}
&& h_{ij}(\vec{x},\eta) = \frac{1}{(2 \pi)^{\frac{3}{2}}}\int d^3 k
\biggl[ q^{(1)}_{ij} h_{\oplus}(\vec{k},\eta)
e^{i \vec{k} \cdot \vec{x}} +
q^{(2)}_{ij} h_{\otimes}(\vec{k},\eta)
 e^{i \vec{k} \cdot \vec{x}}\biggr]~~~,
\nonumber\\
&& \tau^{(T)}_{ij}(\vec{x},\eta) = \frac{1}{(2
\pi)^{\frac{3}{2}}}\int d^3 k
\biggl[ q^{(1)}_{ij} \tau_{\oplus}(\vec{k},\eta)
e^{i \vec{k} \cdot \vec{x}} +
q^{(2)}_{ij} \tau_{\otimes}(\vec{k},\eta)
e^{i \vec{k} \cdot \vec{x}}\biggr]~~~,
\end{eqnarray}
with
\begin{equation}
q^{(1)}_{ij}(\vec{k}) = \frac{1}{2} \biggl[ e^{(1)}_{i}(\vec{k})
e^{(1)}_{j}(\vec{k}) - e^{(2)}_{i}(\vec{k}) e^{(2)}_{j}(\vec{k})
\biggr],~~~
q^{(2)}_{ij}(\vec{k}) = \frac{1}{2} \biggl[ e^{(1)}_{i}(\vec{k})
e^{(2)}_{j}(\vec{k}) + e^{(2)}_{i}(\vec{k})
e^{(1)}_{j}(\vec{k})\biggr].
\end{equation}
Notice that $\vec{k}/|\vec{k}|$, $\vec{e}_{(1)}$, $\vec{e}_{(2)}$ are
three unit vectors orthogonal to each others:
\begin{eqnarray}
&&\frac{\vec{k}}{|\vec{k}|}\equiv \biggl( \hat{k}_{x},\hat{k}_{y},
\hat{k}_{z}\biggr)= \biggl(\sin{\theta} \cos{\phi}, \sin{\theta}
\sin{\phi}, \cos{\theta}\biggr),
\nonumber\\
&&\vec{e}_{(1)} \equiv \biggl(\frac{\hat{k}_{x}}{\sqrt{\hat{k}^2_{x}
+
\hat{k}^2_{y}}}, -\frac{\hat{k}_{y}}{\sqrt{\hat{k}^2_{x} +
\hat{k}^2_{y}}},0\biggr)= \biggl( \sin{\phi}, -\cos{\phi},0 \biggr),
\nonumber\\
&&\vec{e}_{(2)}
\equiv \biggl(\frac{ \hat{k}_{x} \hat{k}_{z}}{\sqrt{\hat{k}^2_{x} +
\hat{k}^2_{y}}},  \frac{ \hat{k}_{y}
\hat{k}_{z}}{\sqrt{\hat{k}^2_{x} +
\hat{k}^2_{y}}},-1 + \hat{k}^2_{z}\biggr) = \pm\biggl( \cos{\theta}
\cos{\phi} , \cos{\theta} \sin{\phi}, - \sin{\theta}\biggr),
\end{eqnarray}
(the $\pm$ in the last formula refers to the cases when $\theta
>\pi/2$ [+] and $\theta<\pi/2$ [-]).
After having extracted the tensor modes from the Fourier space
components of the energy-momentum tensor we can find, with some
algebra, $\tau_{\oplus}(\vec{k})$ and $\tau_{\otimes}(\vec{k})$: 
\begin{eqnarray}
&&\tau_{\oplus}(\vec{k}) = \sqrt{\frac{\pi}{2}} \frac{B_0^2}{\pi^2
L_{s}^4 k^3_{s}} \biggl( \hat{k}_{x}^2 + \hat{k}_{y}^2\biggr) {\cal
Q}(k),~~~\tau_{\otimes}(\vec{k}) = -\sqrt{\frac{\pi}{2}} \frac{B_0^2}{\pi^2
L_{s}^4 k^3_s}  \hat{k}_{x} \hat{k}_{y} \hat{k}_{z}
\biggl( \hat{k}_{x}^2 + \hat{k}_{y}^2 + 2 \hat{k}_{z}^2\biggr){\cal
Q}(k),
\nonumber\\
&&{\cal Q}(k) =\biggl(\frac{7 n^2 - 5}{40}\biggr) \biggl[ 1 
+ \frac{k}{k_{s}} +
{\cal O}\biggl(\biggl(\frac{k}{k_s}\biggr)^2\biggr)\biggr]
e^{-\frac{k}{k_s}}.
\label{pol}
\end{eqnarray}
Then the evolution equations for the two polarizations become
\begin{eqnarray}
&& h_{\oplus}'' + 2 {\cal H} h_{\oplus}' + k^2 h_{\oplus} = - 16 \pi
G
\tau_{\oplus},
\nonumber\\
&&  h_{\otimes}'' + 2 {\cal H} h_{\otimes}' + k^2 h_{\otimes} = - 16
\pi G
\tau_{\otimes}.
\label{poleq}
\end{eqnarray}
We solve the evolution equations for $h_{\oplus}$ and $h_{\otimes}$
by
requiring that at $\eta_1$
\begin{equation}
h_{\oplus}(\eta_1) = h_{\oplus}'(\eta_1) = 0,~~~h_{\otimes}(\eta_1)
= h_{\otimes}'(\eta_1) = 0,
\end{equation}
with the result that
\begin{eqnarray}
h_{\oplus}(\vec{k},\eta) = \frac{16 \pi G}{k^2}
\tau_{\oplus}(\vec{k})
\biggl\{ \frac{ k \eta_1 \cos{[k(\eta- \eta_1)]} + \sin{[k(\eta-
\eta_1)]} - k\eta }{k\eta} \biggr\},
\nonumber\\
h_{\otimes}(\vec{k},\eta) = \frac{16 \pi G}{k^2}
\tau_{\otimes}(\vec{k})
\biggl\{ \frac{ k \eta_1 \cos{[k(\eta- \eta_1)]} + \sin{[k(\eta-
\eta_1)]} - k\eta }{k\eta}\biggr\}.
\end{eqnarray}
Using these expression we can compute the energy density of the
produced gravitational waves. Up to now we were dealing only with one
configuration. In principle after the epoch of electron-positron
annihilation different configurations can be present. If the
configurations are not correlated the polarisations of the produced
gravitational waves will also be stochastically distributed, which
means that
\begin{equation}
\big\langle q^{(s)}_{ij}(\vec{k}) q^{ij}_{(s')}(\vec{k}')\big\rangle=
\delta_{s s'} \delta^{(3)}(\vec{k} - \vec{k}'),
\end{equation}
where $<...>$ denotes a stochastic average.
The total energy density radiated in gravitational waves will then be
\cite{rev1}
\begin{equation}
\rho_{GW}(\eta) =\frac{1}{256 \pi^3 a^2} \int d^3 k \biggl\{
|h_{\oplus}'|^2 + | h_{\otimes}'|^2
+ k^2 \biggl[ |h_{\oplus}|^2 + | h_{\otimes}|^2\biggr]\biggr\},
\label{GW}
\end{equation}
which implies
\begin{equation}
\rho_{GW}(\eta) = \frac{G}{2 \pi} \frac{\eta^2}{a^2} \int \biggl[
\tau_{\oplus}^2 + \tau_{\otimes}^2\biggr] {\cal F}(k\eta) d^3 k,
\label{GW2}
\end{equation}
where
\begin{eqnarray}
{\cal F}(k\eta) &=& \biggl\{ \frac{1}{(k\eta)^6} +
\frac{2}{(k\eta)^4} +
\frac{2}{(k\eta)^2} + \frac{1}{(k\eta)^4} \biggl(
\frac{\eta_1}{\eta}\biggr)^2
+ \frac{2}{(k\eta)^2}
\biggl(\frac{\eta_1}{\eta}\biggr)^2 -  \frac{4}{
(k\eta)^3} \biggl(\frac{\eta_1}{\eta}\biggr) \cos{[ k(\eta- \eta_1)]}
\nonumber\\
&+& \biggl[ \frac{1}{(k\eta)^4} \biggl(\frac{\eta_1}{\eta}\biggr)^2
-
\frac{4}{(k\eta)^3} \frac{\eta_1}{\eta}  -
\frac{1}{(k\eta)^6}\biggr]\cos{[ 2 k(\eta - \eta_1)]}  -
\frac{4}{(k\eta)^3} \sin{[k(\eta- \eta_1)]}
\nonumber\\
&+& \biggl[
\frac{2}{(k\eta)^3} \biggl(\frac{\eta_1}{\eta}\biggr)^2  +
\frac{2}{(k\eta)^5}\biggl(\frac{\eta_{1}}{\eta}\biggr)
 - \frac{2}{(k\eta)^5}\biggr]\sin{[2 k (\eta-
\eta_1)]}\biggr\}.
\end{eqnarray}
Define now the comoving frequency $\omega = k/a$ and recall that
$k\eta = \omega/H$ (where $H= \dot{a}/a$ is the Hubble factor in
cosmic time). By taking $\eta_1$ around the epoch of electron
positron
annihilation (i.e. $T\sim 0.1~{\rm MeV}$) we can compute ${\cal
F}(k\eta)$ at any interesting time $\eta_{0}$ as a function of the
frequency.  Taking $\eta_{0} \simeq \eta_{dec}$ (when $T\simeq 0.26~
{\rm eV}$) the (present) decoupling frequency will be
$\omega_{dec}\sim 10^{-16}$ Hz whereas the frequency corresponding to
the
present horizon will be $\omega_{0} \sim 3.2 \times 10^{-18} h_{100}
{\rm Hz}$. The present frequency corresponding to $T\sim
0.1 ~{\rm Mev}$ is $\omega_{s} \sim 10^{-11}$ Hz and for the
range $\omega_{dec} <\omega<\omega_{s}$ we have that $|{\cal F} (k
\eta_{dec})| =
|{\cal F} (\omega/\omega_{dec})| \sim (\omega/\omega_{dec})^{-2}$.

\renewcommand{\theequation}{5.\arabic{equation}}
\setcounter{equation}{0}
\section{Magnetic Spikes in the gravitational wave spectrum}

For the discussion of  the gravitational wave spectra  produced by
magnetic knots, it is useful to define the  energy spectrum in
critical units,
\begin{equation}
\Omega_{GW}(\omega,t) =\frac{1}{\rho_{crit}} \frac{d \rho_{GW}}{ d
\log{\omega}},
\label{crit}
\end{equation}
which will allow a comparison with the
spectra produced, via gravitational instability, 
in the context of ordinary inflationary
model.
Using Eqs. (\ref{pol}) into Eqs. (\ref{GW}) and (\ref{GW2}) we can
perform the angular integration over the momenta and we are left with
the integration over $d\log{k}$. Passing to the physical frequencies
and using Eq. (\ref{crit}) we get that
\begin{equation}
\Omega_{GW}(\omega,t) \simeq \frac{3}{16 \pi^5}
\frac{799}{4096}\Omega_{\gamma}(t)
\Lambda^2 \biggl(\frac{7 n^2 - 5 }{20}\biggr)^2
\biggl(\frac{\omega}{\omega_s}\biggr)
\biggl(\frac{\omega_{dec}}{\omega_{s}}\biggr)\Biggr\{ 1 + O\Biggl(
\biggl(\frac{\omega}{\omega_s} \biggr)^2\Biggr) \Biggr\}
e^{-\frac{\omega}{\omega_s}},~~~\omega_{dec} <\omega<\omega_{s}
\label{gwknspec}
\end{equation}
(notice that $\Lambda = H_0^2/\rho= {\cal B}_0^2(t)/\rho_{c}(t)$ and
$\Omega_{\gamma}(t)$ is simply the
fraction of critical energy density present in the form of radiation
at a given observation time $t$).
For the purposes of the present Section it is also useful  to
introduce the logarithmic energy spectrum for the knot energy density
\begin{equation}
\Omega_{KN}(\omega,t) = \frac{1}{\rho_{crit}} \frac {d \rho_{KN}}{d
\log{\omega}} ,
\label{KN}
\end{equation}
where $\rho_{KN}(\vec{k})$ is  given by $\tau_{00}(\vec{k})$ after
the
integration over the directions of $\vec{k}$. Using
Eqs. (\ref{fourtens}) and (\ref{limit}) we
get that, in the  range $\omega_{dec}<\omega<\omega_{s}$,
\begin{equation}
\Omega_{KN}(\omega,t) \simeq \frac{1}{\pi^3} \Omega_{\gamma}(t) (n^2
+
1 ) \Lambda \biggl( \frac{\omega}{\omega_{s}}\biggr)^3\biggr[ 1 +
{\cal O}
\biggl(\frac{\omega}{\omega_{s}}\biggr)\biggl]e^{-\frac{\omega}{\omega_s}}.
\label{KN2}
\end{equation}
Obviously, for the consistency of our considerations we have to
impose,
for all the frequencies and for any time, that $\Omega_{KN}(\omega,t) <1$. This simply
means that
the energy density of the knot configurations does not modify the
(radiation dominated) background evolution and can be always treated
as a small perturbation. Given the analytic form of Eq. (\ref{KN2})
we
see that the largest contribution to $\Omega_{KN}(\omega,t)$
comes from modes of
the order of the size of the knot. For higher frequencies the
exponential damping becomes effective. Recalling that, today,
$\Omega_{\gamma}(t_0) \simeq 10^{-4}$, the critical energy
condition $\Omega_{KN}(\omega,t) <1$ implies that, for 
$\omega<\omega_{s}$,
\begin{equation}
\log_{10}{\Lambda} + \log_{10}( n^2 + 1) \leq 4.53.
\label{kncrit}
\end{equation}
It is also worth mentioning that in the above region of parameters
$\Omega_{GW}(\omega,t) <1$ is automatically satisfied.

We want now to compare the logarithmic energy spectrum of the
gravitational waves produced by magnetic knots
 with some (possible) inflationary spectra. It is
 indeed well known that any transition in the curvature inevitably
amplifies the tensor fluctuations of the metric
\cite{gris,star,gris2}. In
particular the transition from a primordial  inflationary
phase to a decelerated one  is associated with the production of a
stochastic background of gravitational waves \cite{infl}, which
decreases as
$\omega^{-2} $ for $\omega_0 <\omega <\omega_{dec}$ and  stays flat
for $\omega_{dec} <\omega< \omega_{dS}$ ( $\omega_{dS} = 10^{11}
\sqrt{H_{dS}/M_{P}} $ Hz is the present frequency associated with the
maximal curvature scale $H_{dS}$ reached during inflation).
The gravitational waves spectrum induced by the inflationary
transition can then be written as
\begin{eqnarray}
&&\Omega_{GW}(\omega,t_0) = \Omega_{\gamma}(t_0)
\biggl(\frac{H_{dS}}{M_{P}}\biggr)^4,~~~~\omega_{dec}<\omega<\omega_{dS},
\nonumber\\
&&\Omega_{GW}(\omega,t_0) = \Omega_{\gamma}(t_0)
\biggl(\frac{H_{dS}}{M_{P}}\biggr)^4
\biggl(\frac{\omega_{dec}}{\omega}\biggr)^2,~~~~\omega_{0}<\omega<\omega_{dec}.
\label{inflspec}
\end{eqnarray}
There are various bounds which should be taken into account. At large
scale the most significant one comes from the anisotropy of the CMBR
\cite{cmbr} which implies that
\begin{equation}
\Omega_{GW}(\omega,t_0) h_{100}^2 < 7\times 10^{-11} , ~~{\rm for}~~
\omega\sim \omega_0.
\label{cmbr}
\end{equation}
For $\omega_{dec} <\omega <\omega_{dS}$, using the bound (\ref{cmbr})
into Eqs. (\ref{inflspec}), we have 
\begin{equation}
\Omega_{GW}(\omega,t_0) \leq 10^{-13},
\label{cond}
\end{equation}
or, in terms of $H_{dS}$ \cite{cmbr},
\begin{equation}
\frac{H_{dS}}
{M_{P}}\leq 10^{-6}.
\end{equation}
At intermediate frequencies $\omega\sim 10^{-8} $ Hz, the extreme
regularity of the pulsar's pulses \cite{tay} imposes
\begin{equation}
\Omega_{GW}(\omega_{P} ,t_{0}) < 10^{-8},~~{\rm for}~~ \omega_{P}
\sim
10^{-8}.
\label{puls}
\end{equation}
Due to the flatness of the logarithmic energy spectrum of Eq.
(\ref{inflspec}) around  $\omega_{P}$, we can clearly see that the
pulsar's bound is easily satisfied if Eq. (\ref{cond}) is satisfied.

\begin{figure}
\centerline{\epsfxsize = 7.5 cm  \epsffile{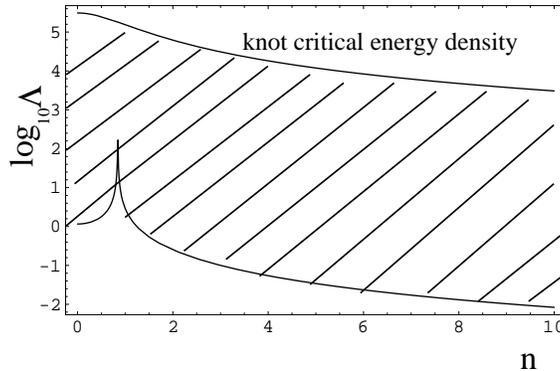}}
\caption[a]{We report the critical energy density bound applied to
the
knot configurations (upper line) and the values of the parameters
$\Lambda$ and $n$ necessary in order to give a signal of the same
order of the stochastic backgrounds of inflationary origin (lower
line). For values of the parameters within the shaded region, the
gravitational wave signal produced by the tensor part of the knot
energy-momentum tensor exceeds the inflationary prediction also of
four orders of magnitude in the (present) frequency
range $\omega_{s}\sim 10^{-11}$--$10^{-12}$ Hz. More
quantitative illustration is given in Fig. \ref{spike}. }
\label{excl}
\end{figure}

We want now to compare the spectrum given in Eq. (\ref{gwknspec})
with
the one produced thanks to the inflationary (parametric)
amplification
of the quantum mechanical fluctuations of the geometry reported in
Eq. (\ref{inflspec}). If we look at Eq. (\ref{gwknspec}) around
$\omega_{s}$ we can see that the gravitational waves produced by the
tensor modes of the knot can exceed $10^{-13}$ (i.e. the maximal
allowed amplitude of the inflationary logarithmic energy spectrum in
critical units) provided:
\begin{equation}
\log_{10}{\Lambda}
+\frac{1}{2}\log_{10}{\biggr[ \biggl(\frac{7 n^2 - 5}{20}\biggr)^2
\biggl] } > -0.537.
\label{bound}
\end{equation}
The last equation was simply obtained  by requiring that
$\Omega_{GW}(\omega, t)$   given in Eq. (\ref{gwknspec}) can be larger than
$10^{-13}$ for $\omega_{dec} <\omega <\omega_{s}$.
In order to be consistent with our approximations we have, therefore
to
impose simultaneously Eqs. (\ref{bound}) and (\ref{kncrit}). The
results are reported in Fig. \ref{excl} in terms of the two parameter
of the model $\Lambda$ (the intensity of the magnetic field in
critical units) and $n$ the number of knots of the configuration.
In Fig. \ref{excl} there are, formally, two parameters. One is
$\Lambda$ and the other is $n$. From the physical point of view,
$\Lambda$ represents the energy scale of the configuration in the
limit
$n\rightarrow 0$, whereas, $n$ is exactly the magnetic helicity
discussed in the previous Sections. We stress the fact that the
helicity is conserved for a wide interval of scales after
$t_{e^{+}e^{-}}$, and, therefore, it represents a very good classical
``label'' which can be used in order to find the gravitational
``imprints'' of  magnetic knots.

\begin{figure}
\centerline{\epsfxsize = 10 cm  \epsffile{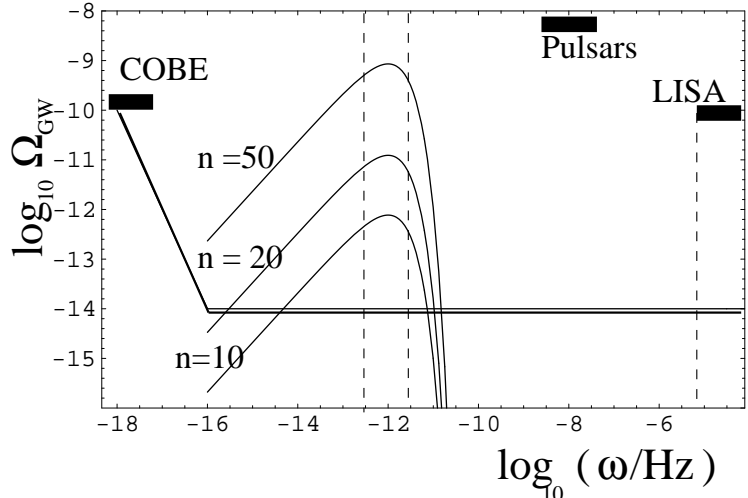}}
\caption[a]{We report various spikes in the gravitational wave
spectrum
computed for different values of $n$. The values of $\Lambda$ is $3$
for the two lower spikes and it is $1$ for the upper one. In this
plot we reported the expected signal computed assuming that quantum
mechanical fluctuations were amplified by the transition between the
inflationary phase and  radiation dominated phase. Notice that the
exponential damping ensures the compatibility with the pulsar bound.}
\label{spike}
\end{figure}
The spectrum
of gravitational waves induced by the knot of Eq. (\ref{gwknspec})
 decreases as $(\omega/\omega_s)$ for $\omega<\omega_s$ and it is
exponentially suppressed for $\omega>\omega_{s}$. This peculiar
behaviour has two important consequences.
Firstly, the exponential suppression guarantees that the pulsar bound
(see Eq. (\ref{puls})) does not constrain the scenario since at that
frequency the gravitational signal of the knot will be suppressed of
a factor of the order of $\exp{[-\omega_{P}/\omega_{s}]}\sim\exp{(-10^4)}$.
Secondly, the steep infrared behaviour guarantees a quick
suppression at scales even larger corresponding to frequencies of the
order of $10^{-16}$ Hz.
Therefore, the signature of these configurations would be compatible
with a spike in the gravitational wave spectrum for frequencies of
the
order of $\omega_{s}$.
Various magnetic spikes are depicted in Fig. \ref{spike}. We can
clearly see that a large magnetic helicity certainly helps in getting
a signal which can be substantially larger (four orders of magnitude
in $\Omega_{GW}$) than the one provided by the stochastic
backgrounds of inflationary origin in the same interval of
frequencies.
In Fig. \ref{spike} we report also the sensitivity of the Laser
Interferometer Space Antenna (LISA) for typical (present) frequencies
of the order of $\omega_{L} \simeq 10^{-3}  -10^{5}$ Hz \cite{danz}.
We would also like to notice that the spikes we just described are
also well below the, so-called, nucleosynthesis bound \cite{nsb}. In
fact the simplest (homogeneous, isotropic) big-bang nucleosynthesis
scenario imposes and (indirect) constraint on the gravitational wave
spectrum which can be written as \cite{revN,revGW}:
\begin{equation}
\int d \log{\omega} \Omega_{GW}(\omega,t_{0}) \leq 0.2~
\Omega_{\gamma}(t_{0}).
\end{equation}
This constraint implies that $\Omega_{GW}(\omega,t_0)$ should be
below $10^{-6}$ as it occurs for all the modes of our spectrum.

Magnetic knots present right after the
epoch of electron-positron annihilation might have, in principle, 
some influence on
the process of galaxy formation. Our discussion will be quite
qualitative and will essentially follow the approach of Ref.
\cite{was}.
Consider the model of a pressure-less fluid made of baryons and of
dark
matter particles with unperturbed velocity fields $u_{B}^{\mu}=
(1/a,0)$, $u_{D}^{\mu} = (1/a,0)$. The velocity fluid perturbations
will then be given, generically by $\delta u^i = v^i/a $. Since in
our
discussion we assumed from the very beginning that $\beta<1$ (as
implied by the critical energy condition on the energy density of the
knot)
the pressure gradients will also be subleading and, therefore the
system for the perturbed density inhomogeneities will be (in the
absence of viscous terms)
\begin{eqnarray}
&&\delta_{B}'' +{\cal H} \delta_{B}' -\frac{3}{2} \biggl[\Omega_{B}
\delta_{B} + \Omega_{D} \delta_{D} \biggr] = \frac{ \vec{\nabla}
\cdot
\bigl[ \vec{H} \times\vec{\nabla} \times \vec{H}\bigr]}{
\bigl[\rho_{D} + \rho_{B}\bigr]},
\label{bdens}\\
&&\delta_{D}'' + {\cal H} \delta_{D} - \frac{3}{2} {\cal H}^2 \biggl[
 \Omega_{B}\delta_{B} + \Omega_{D} \delta_{D} \biggr]=0.
\label{ddens}\\
\end{eqnarray}
where ${\cal H} =a'/a$ is the Hubble factor in conformal time,
$\delta_{B} = \delta_{B}/\rho_{B}$, $\delta_{D} =
\delta\rho_{D}/\rho_{D}$ and $\Omega_{B} = \rho_{B}/[\rho_{B} +
\rho_{D}]$, $\Omega_{D} = \rho_{D}/[\rho_{B} + \rho_{D}]$.
In eq. (\ref{bdens}) the contribution of the Lorentz force has been
written by taking into account the fact the electric field are
suppressed at high conductivity.
Multiplying Eq. (\ref{bdens}) by $\Omega_{B}$ and Eq. (\ref{ddens})
by
$\Omega_{D}$ we find,  after having summed up the two obtained
equations, that the total density contrast $\delta= \Omega_{B}
\delta_{B} + \Omega_{D} \delta_{D}$ obeys a decoupled equation
\begin{equation}
\delta'' + {\cal H} \delta' - \frac{3}{2} {\cal H}^2 \delta =
\Omega_{B}\frac{ \vec{\nabla} \cdot\bigl[ \vec{{\cal B}}
\times\vec{\nabla}
\times \vec{{\cal B}}\bigr]}{\bigl[\rho_{D} + \rho_{B}\bigr]}.
\end{equation}
Notice that, in cosmic time this equation becomes
\begin{equation}
\ddot\delta + 2 \frac{\dot{a}}{a}\dot{\delta} - \frac{3}{2}
\biggl[\frac{\dot{a}}{a}\biggr]^2\delta= \Omega_{B}
\frac{ \vec{\nabla} \cdot\bigl[ \vec{{\cal B}} \times\vec{\nabla}
\times \vec{{\cal B}}\bigr]}{\bigl[a^2(\rho_{D} + \rho_{B}\bigr)]}.
\end{equation}
The solution of this equation in a matter dominated Universe after
recombination ($t>t_{rec}$) can be obtained imposing the initial
conditions $\delta(t_{rec}) =0$ and $\dot\delta(t_{rec})=0$ with the
result that at a given length scale $L$ the density contrast is given
by
\begin{equation}
\delta(L,t) = \frac{144}{15 \pi^3}
\biggl[\frac{\lambda_{B_{J}}(t_{rec})}{L_{s}(t_{rec}}\biggr]^2
\biggl[\frac{a(t)}{a(t_{rec})}\biggr] {\cal G}(\lambda,n)
\label{contrast}
\end{equation}
where $\lambda= L/L_{s}$ and ${\cal G}(\lambda,n)$ contains 
the spatial dependence. In the
case of the configuration given by Eqs. (\ref{knot}) ${\cal
G}(\lambda,n)$ turns out to be:
\begin{eqnarray}
&&{\cal G}(\lambda,n) = P_0(\lambda) + P_{n}(\lambda),
\nonumber\\
&& P_0(\lambda)= \biggl[4 - 19 \lambda^2 + 19 \lambda^4 +
(27 \lambda^2 - 15 \lambda^4) \cos{2 \theta}\biggr],
\nonumber\\
&& P_{n}(\lambda) = \biggl[ 2 \lambda^6  - 16 \lambda^4 + 19
\lambda^2
-5 - \lambda^2\biggl(28 - 13 \lambda^2 + \lambda^4\biggr)\cos{2
\theta}\biggr].
\end{eqnarray}
Notice that ${\cal G}(\lambda,n)$ receives its maximal contribution
around $\lambda=1$. This simply means that the effect of the Lorentz
force term becomes important only for scales comparable with the size
of the knot which, in our scenario turns out to be of the order of
the horizon size right after the epoch of electron-positron
annihilation.

In order to affect structure formation through the Lorentz force we
need that, today \label{was,pee}
\begin{equation}
L_{B_{J}}(t_{0}) \sim 0.03~L_{s}(t_{0})
\label{now}
\end{equation}
with $L_{s}(t_{0}) \sim ~1~{\rm Mpc}$. From this last condition
follows that $\vec{H}(t_{dec}) \sim 10^{-3} {\rm Gauss}$.
Blue-shifting Eq. (\ref{now})
at  the decoupling time we have that
\begin{equation}
L_{B_{J}}(t_{dec}) \sim 2.7\times 10^{19}~{\rm cm}.
\end{equation}
Now,  the typical scale of the knot is $L_{s}(t_{e^{+}e^{-}}) \sim
L_{H}(t_{e^{+}e^{-}})$. This means that $L_{s}(t_{dec}) \sim 10^{18}
{\rm cm}$. Thus we have that, in our scenario,
$L_{B_{J}}\sim 10 ~L_{s}(t_{dec})$.
As we just said, in order to have a significant impact on structure
formation, we would need $L_{s}(t_{dec})\sim L_{s}(t_{dec})$,
whereas, in our case, we have $L_{s}(t_{dec}) \sim
0.1~L_{s}(t_{dec})$ which might have some mild impact on structure
formation. In the spirit of the order of magnitude estimate presented
in the last part of the present Section we notice that, if
appreciable effects occur on structure formation (i.e.
$|\vec{H}(t_{dec})|\sim 10^{-3} {\rm Gauss}$) the very same amplitude
of the magnetic field can depolarise the CMBR \cite{pol}.

\renewcommand{\theequation}{5.\arabic{equation}}
\setcounter{equation}{0}
\section{Conclusions}

The dynamical symmetries of hot plasmas impose the conservation of
the
magnetic flux and of the magnetic helicity. In this paper we gave
some
examples of magnetohydrodynamical configurations carrying a finite
magnetic helicity. We argued that, thanks
to the helicity conservation the topological properties of the
divergence free magnetic field lines are the only dynamically
motivated parameter which could be used in order to discuss the
possible implications of magnetic fields in the early Universe.

Based on the fact that after the epoch of electron-positron
annihilation the MHD are fully valid we argued that if magnetic knots
were present they could have acted as a source for the
evolution equation of the scalar, vector and tensor modes of the
metric inhomogeneities. We focused our attention on the tensor modes
and we found that the induced gravitational wave spectrum is sharply
peaked around a (present) frequency of the order of $10^{-11}$
-$10^{-12}$ Hz. The typical amplitude of the spike depends upon the
helicity of the configuration or, more physically, upon the
the
number of twists and knots in the magnetic flux lines. The amplitude
of the spikes can be substantially larger than the gravitational
waves
 signal provided, in the same interval of frequencies, by
inflationary models.

Many interesting points are left for future discussions. First of all
there are technical points which could be addressed. In this paper we
only took into account the tensor modes. Of course it is certainly
possible to extend the above discussion to scalar and vector
modes. It is not excluded that some interesting consequences for the
small angular scales anisotropies can be addressed. Moreover we would
like to stress that the present results hold in the case of one
configuration. It would be interesting to generalise these results to
more complicated statistical distributions of magnetic knots. 
This problem has not been addressed in our present investigation.

There are also deeper questions. In this investigation we were
dealing with ordinary MHD. It is, in principle possible to extend our
discussion to earlier times (like the electroweak epoch) \cite{mis1,mis3}.

\section*{Aknowledgments} 

I would like to thank M. Shaposhnikov for many useful discussions.
\newpage

\begin{appendix}
\renewcommand{\theequation}{A.\arabic{equation}}
\setcounter{equation}{0}
\section{Magnetic flux and helicity theorems}

Consider and arbitrary closed surface $\Sigma$ which moves with the
plasma. Then, by definition of the bulk velocity of the plasma
($\vec{v}$ we can also write  $d\vec{\Sigma} = \vec{v} \times
d\vec{l}~d\eta$. The (total) time derivative of the flux can
therefore
be expressed as
\begin{equation}
\frac{d}{d\eta} \int_{\Sigma} \vec{H}\cdot d\vec{\Sigma}=
\int_{\Sigma} \frac{\partial \vec{H}}{\partial\eta} \cdot
d\vec{\Sigma} + \int_{\partial\Sigma} \vec{H}\times\vec{v} \cdot
d\vec{l}
\label{first}
\end{equation}
where $\partial \Sigma$ is the boundary of $\Sigma$. Using now the
properties of the vector products
(i.e. $\vec{H}\times\vec{v}\cdot d\vec{l}= - \vec{v}\times\vec{H} \cdot
d\vec{l} $) we can express $\vec{v} \times \vec{H}$ though the Ohm
law
given in Eq. (\ref{ohm}) and we obtain that
\begin{equation}
\vec{v} \times \vec{H} = - \vec{E} + \frac{1}{\sigma} \vec{\nabla}
\times \vec{H}
\end{equation}
Using now Eq. (\ref{first}) together with the Stokes
theorem we obtain
\begin{equation}
\frac{d}{d\eta} \int_{\Sigma} \vec{H}\cdot d\vec{\Sigma}=\int_{\Sigma} \biggl[
\frac{\partial \vec{H}}{\partial\eta} + \vec{\nabla} \times
\vec{E}\biggr]\cdot d\vec{\Sigma} - \frac{1}{\sigma}
\int_{\Sigma} \vec{\nabla} \times\vec{\nabla}
\times\vec{H}\cdot d\vec{\Sigma}
\label{second}
\end{equation}
From Eq. (\ref{Mx1}) given in Sec. II the first part at the right
hand side of Eq. (\ref{second}) is zero and Eq. (\ref{flux}),
expressing the
Alfv\'en theorem, is recovered.

With a similar technique we can show that the conservation of the
magnetic helicity also holds. Consider a closed volume in the plasma,
then we can write that $dV = d^3 x= \vec{v}_{\perp}\cdot
d\vec{\Sigma}~
d\eta \equiv \vec{n} \cdot \vec{v}_{\perp}~ d\Sigma~ d\eta$ where
$\vec{n}$ is the unit vector normal to $\Sigma$ (the boundary of V, i.e.
$\Sigma = \partial V$) and $\vec{v}_{\perp}$ is the component of the
bulk velocity orthogonal to $\partial V$. The (total) time derivative
of the magnetic helicity can now be written as
\begin{equation}
\frac{d}{d\eta} {\cal H}_{M} = \int_{V} d^3 x
\frac{\partial}{\partial\eta} \bigl[\vec{A} \cdot\vec{H}\bigr] +
\int_{\partial V=\Sigma}\vec{A}\cdot\vec{H}
\vec{v}_{\perp}\cdot\vec{n}
d\Sigma
\end{equation}
By now writing down explicitly the partial derivative at the right
hand side of the previous equation we can use the MHD equations
(recall that $\vec{E}= - \vec{A}'$ and that
$\vec{H} =\vec{\nabla}\times\vec{A}$). Finally using again the Ohm
law and transforming the obtained surface integrals into volume
integrals (through the divergence theorem) we get to Eq. (\ref{h2}).

We want to stress the fact the the magnetic
helicity is indeed a gauge invariant quantity under the conditions
stated in Sec. II. Consider a gauge transformation
\begin{equation}
\vec{A} \rightarrow \vec{A}  + \vec{\nabla} \xi
\end{equation}
then the magnetic helicity changes as
\begin{equation}
\int_{V} d^3x \vec{A} \cdot \vec{H} \rightarrow
\int_{V} d^3x \vec{A} \cdot \vec{H} + \int_{V} d^3 x
\vec{\nabla}\cdot\bigl[
\xi \vec{H}\bigr]
\end{equation}
(in the second term at the right hand side we used the fact that the
magnetic field is divergence free). By now using the divergence
theorem we can express the volume integral as
\begin{equation}
 \int_{V} d^3 x  \vec{\nabla}\cdot\bigl[ \xi \vec{H}\bigr] =
 \int_{\partial V=\Sigma } \xi\vec{H} \cdot\vec{n} d\Sigma
\end{equation}
Now if, as we required, $\vec{H}\cdot\vec{n}=0$ in $\partial V$, the
integral is exactly zero and ${\cal H}_{M} $ is gauge invariant.
Since
in Sec. II we considered fields going  to zero at infinity
$\vec{H}=0$ in any part of $\partial V$ at infinity. Notice that the
field go to zero at infinity is not really necessary. In fact we
could
just define the integration volume by slicing $V$ in a collection  of
closed
 flux tubes. Now by definition of flux tube $\vec{H}\cdot\vec{n}=0$
if
$\vec{n}$ is the unit vector orthogonal to the flux tube.

In order to give a self-contained presentation of our problem we want
to recall the main limits where the MHD equations can be
discussed. This part of the Appendix represents an ideal follow up of
the considerations presented in Section II.

Using the Ohm law of Eq. (\ref{ohm}) in order to express the induced
electric field in terms of the current density  and taking
into account that $\vec{J} = \vec{\nabla}\times\vec{H}$, we obtain
from
 Eq. (\ref{Mx2}) the standard form of the magnetic diffusivity
equation
\begin{equation}
\vec{H}' = \vec{\nabla}\times\bigl[ \vec{v} \times\vec{H}\bigr] +
\frac{1}{\sigma} \nabla^2 \vec{H} ,~~~\vec{\nabla}\cdot H= 0.
\label{diff}
\end{equation}
Assuming the incompressible closure together with the
adiabatic assumption (i.e. $\rho'=0,~p'=0$) we get a simpler form of
the Navier Stokes equation
\begin{equation}
\vec{v}~' + \bigl[\vec{v}\cdot\vec{\nabla}\bigr] \vec{v}
= - \frac{1}{p
+ \rho} \vec{\nabla} \biggl[ p + \frac{|\vec{H}|^2}{2}\biggr] +
\bigl[
\vec{H}\cdot\vec{\nabla}\bigr] \vec{H} + \frac{\nu}{p
+\rho}\nabla^2\vec{v}.
\label{ns2}
\end{equation}
In writing this equation we used the fact that $2~\vec{J}
\times\vec{H} = -\vec{\nabla}|\vec{H}|^2 + 2
[\vec{H}\cdot\vec{\nabla}]\vec{H}$.
 Eqs. (\ref{diff}) and (\ref{ns2}) contain  
the magnetic   and  thermal diffusivity scales. We will
 focus our attention on the case of an electromagnetic plasma whose
 temperature is $T\ll 100~{\rm GeV}$ (in the opposite
 limit [$ T> 100~ {\rm GeV}$] the
 equations appearing in the present Section should indeed be
 generalized to the case of finite chemical potential leading to the
 anomalous MHD equations \cite{mis1}).
An estimate of the plasma conductivity is necessary for the calculation
of  the magnetic diffusivity scale. Before the epoch of
 $e^{+}$--$e^{-}$ annihilation (i.e. $T> 2 m_{e}\sim 1 ~ {\rm MeV}$)
 the conductivity can be written as
$\sigma_{c} = \lambda T/{\alpha_{em}}$ where $\lambda$ is a constant
 of order unity (recall also that, from Eq. (\ref{defin}),
 $\sigma(\eta) = \sigma_{c}~a(\eta)$). For temperatures smaller than
 the weak interactions equilibration temperature (i. e. $T< 0.2
 ~m_{e}$) the electron velocity can be roughly given as
$v = \sqrt{T/m_{e}}$ whereas the collision cross-section is
  $\sigma_{coll} \simeq (\alpha_{em}/T)^2$ (up to the
 Coulomb logarithm which is of order one in our case). Therefore the
conductivity will be simply estimated by
\begin{equation}
\sigma_{c} \simeq \frac{\alpha_{em}}{m_{e}} \frac{1}{\sigma_{coll}
v}\simeq \frac{1}{\alpha_{em}}\sqrt{\frac{T^3}{m_{e}}}.
\label{conduc}
\end{equation}
From Eq. (\ref{diff}) the magnetic diffusivity scale will be 
\begin{equation}
L_{\sigma}(\eta) \sim \sqrt{ \frac{\eta}{\sigma_{c} a(\eta)}}=
\sqrt{\frac{t}{\sigma_{c}}}.
\end{equation}
Using Eq. (\ref{conduc}) we can say that, after the weak decoupling,
\begin{equation}
L_{\sigma}(T)\simeq 4.3~ \times~ 10^{10} ~
\biggl(\frac{T}{0.308 {\rm eV}}\biggr)^{- \frac{7}{4}} ~{\rm
cm},~~~T< ~0.2~m_{e}
\label{diffscale}
\end{equation}
(where we used as reference the recombination temperature 
 $T_{rec}\sim 0.308~{\rm eV}$). Notice that this scale is
incredibly small if compared with the horizon size at each
corresponding epoch.
The magnetic diffusivity scale tells us which is the
``inertial'' \cite{bis,ol} range of the magnetic field spectrum
where the effect of the finite value of the conductivity can be
approximately neglected (the so called ideal limit of the MHD
equations).
The thermal diffusivity coefficient is given by
$\nu/(p+ \rho) \sim 0.2~ \lambda_{\gamma} (T)$ where
\begin{equation}
 \lambda_{\gamma}(T) \sim 9.106 \times
 10^{10}~x^{-1}_{e}~(\Omega_{B}~h^2_{100})^{-1}( {\rm MeV}/T)^2
T^{-1}
\end{equation}
 ( $x_{e}$ is the
 ionization fraction and $\Omega_{B}~ h^2_{100}$ is the fraction of
critical energy density of baryons). Therefore the thermal
diffusivity scale will be
\begin{equation}
 L_{th}(T) \sim 1.5\times 10^{16} x^{-\frac{1}{2}}_{e} (\Omega_{B}
h^2_{100})^{-\frac{1}{2}}\biggl(\frac{ {\rm
MeV}}{T}\biggr)^{\frac{3}{2}}~T^{-1}.
\end{equation}
It is important to notice that the ratio of these two quantities
gives
essentially the (magnetic) Prandtl number \cite{bis}
\begin{equation}
Pr_{m}(T) = 3.5 \times 10^{12} ~x_{e}^{-1} (\Omega_{B}h^2_{100})^{-1}
\biggl( \frac{\rm MeV}{T}\biggr)^{\frac{3}{2}} .
\end{equation}
Since  $Pr_{m}(T_{dec}) \gg 1$ the effect of the Lorentz force terms
in Eq. (\ref{ns2}) cannot be neglected. Another interesting
dimensionless ratio is what is usually called $\beta$-parameter \cite{bis}
\begin{equation}
\beta(T)= \frac{ |\vec{H}(T)|^{2}}{2~p}\sim \frac{45}{2 \pi^2
g_{eff}}
 \frac{|\vec{H}(T)|^2}{T^4}
\end{equation}
If $\beta\ll 1$, then the term $\vec{\nabla}[p + |\vec{H}|^2/2]$ can
be neglected. The critical energy condition applied on the magnetic
field backgrounds at any given temperature in the radiation dominated
epoch enforces by itself this assumption. In order to give a more
quantitative estimate, let us suppose that  the magnetic field is
strong enough to produce an
appreciable depolarisation of the CMBR. If this is the case we should
have $|\vec{H}(t_{dec})| \sim
10^{-3}$ Gauss \cite{pol} around $T_{dec} \sim 0.26$ eV. This means
that  $\beta(T_{dec}) \sim 2.7 ~10^{-2}$.
Therefore for large $Pr_{m}$ and small $\beta$
Eqs. (\ref{diff})--(\ref{ns2}) can be further simplified:
\begin{eqnarray}
&&\vec{v}~' + \bigl[\vec{v}\cdot\vec{\nabla}\bigr]\vec{v}
= \frac{1}{p + \rho} \bigl[\vec{H} \cdot\vec{\nabla}\bigr]\vec{H},
\label{ns3}\\
&& \vec{H}' = \vec{\nabla}\times\bigl[ \vec{v} \times\vec{H}\bigr].
\end{eqnarray}

\renewcommand{\theequation}{B.\arabic{equation}}
\setcounter{equation}{0}
\section{Energy-momentum tensor for knot configurations}

In this appendix we want to give the components of the energy
momentum
tensor which are used in our calculations. We want to express the
energy-momentum tensor in Cartesian coordinates where some estimates
become more tractable. Let us start by giving the cartesian components
of the magnetic field:
\begin{eqnarray}
H_{x}(\vec{{\cal X}}) &=& \frac{4 B_0}{\pi L_{s}^2}\frac{ 2 {\cal
 Y} - 2~ n~
{\cal X}{\cal Z}}{\bigl[1 + {\cal X}^2 + {\cal Y}^2 + {\cal
Z}^2\bigr]^3},
\nonumber\\
H_{y}(\vec{{\cal X}}) &=& -
\frac{4 B_0}{\pi L_{s}^2}\frac{ 2 {\cal X} + 2~n~
{\cal Y} {\cal Z}}{\bigl[1 + {\cal X}^2 + {\cal Y}^2 + {\cal
Z}^2\bigr]^3},
\nonumber\\
H_{z}(\vec{{\cal X}}) &=&
  \frac{4 B_0}{\pi L_{s}^2} \frac{ n \bigl[ {\cal
X}^2 + {\cal Y}^2 - {\cal Z}^2 -1\bigr]}{\bigl[1 + {\cal X}^2 + {\cal
Y}^2 + {\cal Z}^2\bigr]^3},
\label{cart}
\end{eqnarray}
where, as usual, ${\cal X} = x/L_{s}.~{\cal Y} = y/L_{s},~{\cal Z}=
z/L_{s}$.
The spherical components reported in Eq. (\ref{knot}) are simply related
to the Cartesian ones.
In Cartesian coordinates we have
\begin{equation}
\vec{H}(\vec{{\cal X}}) = H_x \vec{e}_{x} + H_{y} \vec{e}_{y} +
H_{z} \vec{e}_{z}
\label{vecc}
\end{equation}
where $\vec{e}_{x}$, $\vec{e}_{y}$ and $\vec{e}_{z}$ are three
mutually orthogonal unit vectors. The three unit vectors in spherical
polar coordinates are  determined by computing the Jacobian
matrix with the result that:
\begin{eqnarray}
\vec{e}_{r} &=& \frac{{\cal X}}{{\cal R}} \vec{e}_{x} +\frac{{\cal
Y}}{{\cal R} }\vec{e}_{y} +  \frac{{\cal Z}}{{\cal R}} \vec{e}_{z},
\nonumber\\
\vec{e}_{\theta} &=& \frac{ {\cal Z} {\cal X}}{{\cal R} \sqrt{ {\cal
R}^2 - {\cal Z}^2}} \vec{e}_{x} +
\frac{ {\cal Z} {\cal Y}}{{\cal R} \sqrt{ {\cal
R}^2 - {\cal Z}^2}} \vec{e}_{y} - \frac{\sqrt{ {\cal R}^2 - {\cal
Z}^2
}}{{\cal R}^2}\vec{e}_{z},
\nonumber\\
\vec{e}_{\phi} &=& -\frac{{\cal Y}}{\sqrt{{\cal X}^2 + {\cal
Y}^2}}\vec{e}_{x} + \frac{{\cal X}}{\sqrt{{\cal X}^2 + {\cal
Y}^2}}\vec{e}_{y}.
\label{trans}
\end{eqnarray}
The polar expression for the  components of the magnetic fields
 are obtained, from Eqs. (\ref{cart}),
(\ref{vecc}) and (\ref{trans}) as $H_{r} = \vec{H} \cdot
\vec{e}_{r}$, $H_{\theta} = \vec{H} \cdot \vec{e}_{\theta}$,
$H_{\phi}
= \vec{H}\cdot\vec{e}_{\phi}$. In order to get to recover the result
 of Eq. (\ref{knot}) the well known relations between spherical and
 Cartesian coordinates should be used (i.e. ${\cal X} = {\cal R}
 \sin{\theta} \sin{\phi}$, ${\cal Y} = {\cal R} \sin{\theta}
 \cos{\phi}$, ${\cal Z} = {\cal R} \cos{\theta}$).

Recall now that $\vec{H}(\vec{{\cal X}}) = a^2 \vec{\cal B}$. In
terms
of $\vec{{\cal B}}$ the Maxwell field strength is given by $F_{ij} =
a^2\epsilon_{ijk}{\cal B}^{k}$ and the corresponding energy-momentum
tensor will be
\begin{equation}
\delta T_{\mu\nu} = \biggl( F_{\mu\alpha} F^{\alpha}_{\nu} -
\frac{1}{4} g_{\mu\nu} F_{\alpha\beta} F^{\alpha\beta} \biggr)
\end{equation}
(we kept the notation $\delta T_{\mu\nu}$ used in Section 4 in order
to stress the fact that $\vec{\cal B}$ is nothing but a fluctuation
on
the homogeneous background).
It is often useful to express the components of the energy momentum
tensor (as we did) in terms of the rescaled fields $\vec{H}$. The
relation between the energy-momentum tensors  is
$\delta T_{\mu\nu}=a^{-2} \tau_{\mu\nu}$
($\delta T^{\mu\nu}= a^{-6} \tau^{\mu\nu}$),
where now $ \tau_{\mu\nu}$ is only expressed in terms of $\vec{H}$
(remember that we work always in conformal time ).
The components of the Energy-momentum tensor are:
\begin{eqnarray}
\tau_{0 0}(\vec{\cal X}) &=& \frac{8 B_0^2}{L_{s}^4 \pi^2} \frac{ 4
\bigl( {\cal X}^2
+ {\cal Y}^2 \bigr) + n^2 \bigl[ 1 + {\cal X}^4 - 2 {\cal Y}^2 +
{\cal
Y}^4 + 2 {\cal Z}^2  + 2 {\cal Y}^2 {\cal Z}^2 + {\cal Z}^4 + 2 {\cal
X}^2 \bigl( {\cal Y}^2 + {\cal Z}^2 -1 \bigr)\bigr]}{\bigl[1 + {\cal
X}^2 +
{\cal Y}^2 + {\cal Z}^2\bigr]^6},
\nonumber\\
\tau_{x x}(\vec{\cal X}) &=&   \frac{8
B_0^2}{L_{s}^4\pi^2}\frac{4\bigl( {\cal X}^2 -
{\cal Y}^2\bigr) + 16~n {\cal X}{\cal Y}{\cal Z} + n^2 \bigl[ 1 +
{\cal X}^4 - 2 {\cal Y}^2 + {\cal Y}^4 + 2 {\cal Z}^2 + 2 {\cal Z}^2
{\cal Y}^2 + {\cal Z}^4 + 2 {\cal X}^2 \bigl( {\cal Y}^2 -
3 {\cal Z}^2 -1\bigr)\bigr]}{\bigl[1 + {\cal X}^2
 +{\cal Y}^2 + {\cal Z}^2\bigr]^6},
\nonumber\\
\tau_{y y}(\vec{\cal X}) &=& - \frac{8 B_0^2}{L_{s}^4 \pi^2} \frac{4
\bigl( {\cal X}^2 -
{\cal Y}^2\bigr) + 16 {\cal X}{\cal Y} {\cal Z} - n^2 \bigl[ 1 +
{\cal X}^4
- 2 {\cal Y}^2 + {\cal Y}^4 + 2 {\cal Z}^2 - 6 {\cal Y}^2 {\cal Z}^2
+
{\cal Z}^4 + 2 {\cal X}^2 \bigl( {\cal Y}^2 + {\cal Z}^2
-1\bigr)\bigr]}{\bigl[1 + {\cal X}^2 +{\cal Y}^2 + {\cal
Z}^2\bigr]^6}
\nonumber\\
\tau_{z z}(\vec{\cal X}) &=& -\frac{8 B_0^2}{L_{s}^4 \pi^2}\frac{-
4\bigl( {\cal X}^2
+ {\cal Y}^2\bigr) + n^2 \bigl[ 1 + {\cal X}^4 - 2 {\cal Y}^2 + {\cal
Y}^4 + 2 {\cal Z}^2 - 6 {\cal Y}^2 {\cal Z}^2 + {\cal Z}^4 + 2 {\cal
X}^2 \bigl( {\cal Y}^2 - 3 {\cal Z}^2 -1\bigr)\bigr]}
{\bigl[1 + {\cal X}^2 +{\cal Y}^2 + {\cal Z}^2\bigr]^6},
\nonumber\\
\tau_{x y}(\vec{\cal X}) &=&  \frac{64 B_0^2}{L_{s}^4 \pi^2}
\frac{\bigl({\cal Y} -
n{\cal X} {\cal Z}\bigr)\bigl({\cal X} + n {\cal Y} {\cal Z}\bigr)}
{\bigl[1 + {\cal X}^2 +{\cal Y}^2 + {\cal Z}^2\bigr]^6},~~~
\tau_{x z}(\vec{\cal{X}}) = \frac{32 B_0^2 n}{L_{s}^4 \pi^2}
\frac{\bigl( - {\cal Y} +
n{\cal X}{\cal Z}\bigr) \bigl( {\cal X}^2 +{\cal Y}^2 - {\cal Z}^2
-1\bigr)}
{\bigl[1 + {\cal X}^2 +{\cal Y}^2 + {\cal Z}^2\bigr]^6},
\nonumber\\
\tau_{y z}(\vec{\cal X}) &=&  \frac{32 B_0^2 n}{L_{s}^4  \pi^2
}\frac{ \bigl( {\cal
X} + n{\cal Y} {\cal Z}\bigr) \bigl( {\cal X}^2
+{\cal Y}^2 - {\cal Z}^2 -1\bigr)}
{\bigl[1 + {\cal X}^2 +{\cal Y}^2 + {\cal Z}^2\bigr]^6}.
\label{enmom}
\end{eqnarray}
Concerning this energy-momentum tensor few comments are in order. We
did not include any electric field. The reason is simply that all the
parts of the energy-momentum tensor containing electric parts are
sub-leading at high conductivity. In fact, as we discussed in Section
 II in our approximations the electric fields are given by $\vec{E}
\sim \vec{\nabla}\times\vec{H} /\sigma$ and, therefore, negligible in
the ideal approximation of MHD. However, if one would like to study
 precisely the border region of the ideal approximation
(i.e. typical scales of the knot comparable with the magnetic
diffusivity scale given by Eq. (\ref{diffscale})), then, the electric
components should be included (sometimes this approximation scheme is
named resistive MHD \cite{bis}). In our
discussion the knot scale $L_{s}$ is comparable with the magnetic
Jeans length and then much larger than the magnetic diffusivity
scale.
There are interesting limits where the various components of
$T_{\mu}^{\nu}$ could be investigated like
$n\rightarrow 0$.

\renewcommand{\theequation}{C.\arabic{equation}}
\setcounter{equation}{0}
\section{Fourier transforms of the Energy-momentum tensor components}

In this Appendix we report the Fourier transforms of the components
of
the energy momentum tensor of our configurations which are crucial
for
the discussion of the induced inhomogeneities.
\begin{eqnarray}
\tau_{xx}(\vec{k}) &=& \frac{8 B_0^2 }{\pi^2 L_{s}^4 k^3_{s}}
\biggl\{
n^2 \biggl[ I_{6}(\kappa)  - \frac{1}{5} I_{5}(\kappa) \biggr] +
\frac{1}{80}I_{4}(\kappa) \biggl[ 2 (n^2 -2) \kappa_{x}^2
+ 2(n^2 + 2) {\kappa}_{y}^2 - 2 n^2 {\kappa}_{z}^2 + 7 n^2\biggr]
\nonumber\\
&+&
\frac{n}{240} I_{3} (\kappa) \biggl[ - n {\kappa}^2_{x}- n ( 5
\kappa^2_{y} + \kappa^2_{z} + 8 i {\kappa}_{x} {\kappa}_{y}
{\kappa}_{z}\biggr] + \frac{n^2}{1920} I_{2}(\kappa) \biggl[
{\kappa}^4_{x} + 2 {\kappa}^2_{x}(\kappa^2_{y} - 3 {\kappa}^2_{z}) +
(\kappa^2_{y} + {\kappa}^2_{z})^2\biggr]\biggr\},
\nonumber\\
\tau_{yy}(\vec{k}) &=& \frac{ 8 B^2_{0} }{ \pi^2 L^4_{s}k^3_{s}}
\biggl\{ n^2 \biggl[ I_{6} (\kappa)  - \frac{1}{5} I_{5}(\kappa)
\biggr] + \frac{1}{80} I_{4}(\kappa)\biggl[ 2 (n^2 + 2) \kappa^2_{x}
+
2 (n^2 - 2)\kappa^2_{y} - 2 n^2 \kappa^2_{z} + 7 n^2 \biggr]
\nonumber\\
&+&\frac{n}{240} I_{3}(\kappa) \biggl[ ( 2 {\kappa}^2_{y} + 2
{\kappa}^2_{z} - 5 \kappa^2_{x}) - 8 i  {\kappa}_{x} \kappa_{y}
\kappa_{z}\biggr] + \frac{n^2}{5760} I_{2}(\kappa)
\biggl[\kappa^4_{x} + 2
{\kappa}^2_{x}( \kappa^2_{y} + \kappa^2_{z}) + ( \kappa^4_{y} - 6
\kappa^2_{y} \kappa^2_{z}+ \kappa^4_{z})\biggr]\biggr\},
\nonumber\\
\tau_{zz}(\vec{k}) &=& \frac{ 8 B^2_{0} }{ \pi^2
L^4_{s} k^3_s}\biggl\{ - n^2 I_{6}(\kappa) + \frac{ n^2 + 4}{5}
I_{5}(\kappa) + \frac{1}{80} I_{4}(\kappa) \biggl[ 2 n^2 \kappa^2_{z}
- 2 (n^2 + 2) \kappa_{x} - 2(n^2 +2)\kappa_{y}\biggr]
\nonumber\\
&+& \frac{n^2}{240} I_{2}(\kappa)\biggl[ \kappa^2_{x} - 2
(\kappa^2_{y} + 3 \kappa^2_{z})\biggr] - \frac{n^2}{1920}
I_{2}(\kappa) \biggl[ \kappa^4_{x} + \kappa^4_{z} + \kappa^2_{y} - 6
\kappa^2_{y}\kappa^2_{z} + 2 \kappa^2_{x}( \kappa^2_{y} -3
\kappa^2_{z})\biggr]\biggr\},
\nonumber\\
\tau_{xy}(\vec{k}) &=&  \frac{ 64 B^2_{0} }{ \pi^2 L^4_{s} k^3_{s}}
\biggl\{ - \frac{n^2}{1920} I_{2}(\kappa) \kappa_{x} \kappa_{y}
\kappa^2_{z} - \frac{1}{80} I_{4}(\kappa) \kappa_{x} \kappa_{y}  +
\frac{n}{480} I_{2}(\kappa) \biggl[ \kappa_{x} \kappa_{y} + i(
\kappa^2_{y}\kappa_{z} - \kappa^2_{x}\kappa_{z})\biggr]\biggr\},
\nonumber\\
\tau_{xz}(\vec{k}) &=&   \frac{ 32 B^2_{0} }{ \pi^2 L^4_{s}k^3_{s}} n
\biggl\{ \frac{1}{1920} I_{2}(\kappa) \biggl[ \kappa_{x} \kappa_{z} (
\kappa^2_{x} - \kappa^2_{z} +n \kappa^2_{y})\biggr] -
\frac{i}{480}
I_{3}(\kappa) \biggl[ \kappa^2_{x}\kappa_{y} + \kappa^3_{y} - 8
\kappa_{z} - \kappa_{y} \kappa^2_{z} + \kappa_{x} ( 8 - i n
\kappa_z)\biggr]
\nonumber\\
&+& \frac{1}{160} I_{4}(\kappa) \biggl[ 2 n \kappa_{x} \kappa_{z} - 3
i
\kappa_{y}\biggr] - \frac{i}{10} I_{5}(\kappa) \kappa_{y}\biggr\},
\nonumber\\
\tau_{yz}(\vec{k}) &=&  \frac{ 32 B^2_{0} }{ \pi^2 L^4_{s} k^3_{s}}
\biggl\{ \frac{1}{1920} I_{2}(\kappa) \biggl[ \kappa_{y} \kappa_{z} (
\kappa^2_{y} - \kappa^2_{z} + n\kappa^2_{x})\biggr] + \frac{i}{480} 
I_{3}(\kappa)
\biggl[  \kappa^3_{x} - 8 ( \kappa_{y} - \kappa_{z} ) + \kappa_{x}
( 48 + \kappa^2_{y} - \kappa^2_{z} + i n \kappa_{z})\biggr]
\nonumber\\
&+& \frac{1}{160} I_{4}(\kappa) \biggl[ 3 i \kappa_{x} + 2 n
\kappa_{y}
\kappa_{z}\biggr]\biggr\},
\nonumber\\
\tau_{00}(\vec{k}) &=&  \frac{ 8 B^2_{0} }{ \pi^2 L^4_{s} k^3_s}
\biggl\{ \frac{1}{1920} I_{2}(\kappa)\biggl[ n^2 ( \kappa^4_{x} +
\kappa^4_{y} + \kappa^4_{z}) + 2 n^2 \kappa^2_{z}\kappa^2_{y}
+ 2  n^2 \kappa^2_{x} \kappa^2_{y} + 2
\kappa^2_{x}\kappa_{z}^2\biggr]
+ \frac{4 - n^2}{5}
I_{5}(\kappa)  + n^2 I_{6}(\kappa)
\nonumber\\
&-&  \frac{1}{240} I_{3}(\kappa)
\biggl[ 3 n^2 ( \kappa^4_x + \kappa^4_{y} + \kappa^4_{z}) + n^2 (
\kappa^2_{x} + \kappa^2_{y} + 2 \kappa^2_{z})\biggr]
\nonumber\\
&-& \frac{1}{80}
I_{4}(\kappa) \biggl[ 15 n^2 - 2 n^2 ( \kappa^2_{z} + \kappa_{x}^2 -
\kappa^2_{y}) - 4 (\kappa^2_{x} + \kappa^2_{y})\biggr] \biggr\}.
\label{fourtens}
\end{eqnarray}
Notice that $\kappa_{i} \equiv k_{i} /k_{s}$ and that $\kappa =
 \sqrt{\kappa_{i}\kappa^{i}}$. In the same
way as we reported the Cartesian components of the energy momentum
tensor  in terms of the
``rescaled'' coordinates $({\cal X}, {\cal Y}, {\cal Z}) =
(x,y,z)/L_{s}$, also dealing with the corresponding Fourier
components we find useful to work with dimensionless momenta. The
fact that we work with $\kappa $ (instead of with $k$) also implies
that a factor $k^{-3}_s$ appears in front of every component of the
Fourier components of the energy-momentum tensor.
Concerning Eqs. (\ref{fourtens}) different comments are in
order. First of all we can notice that at small scales (i.e. $\kappa
>1$) the Fourier components of the energy momentum tensor are
exponentially suppressed as $\exp{[-\kappa]}= \exp{[-k/k_{s}]}$. In
other words the typical scale of the knot acts as ultraviolet
cut-off. At large scales (i.e. for the infrared branch of the
spectrum
$k< k_{s}$) it is of some interest to expand the obtained expressions
and we find that
\begin{eqnarray}
&&\tau_{00}(\vec{k})= \frac{ B_0^2 }{\pi^2
L_{s}^4 k_s^3} \sqrt{\frac{\pi}{2}}\frac{n^2+1}{4} \biggl[ 1 + \kappa
+ {\cal
O}(\kappa^2) \biggr],
\nonumber\\
&&\tau_{xx}(\vec{k}) \sim \tau_{yy}(\vec{k}) = \frac{ B_0^2 }{\pi^2
L_{s}^4 k_s^3} \sqrt{\frac{\pi}{2}}\frac{n^2}{5} \biggl[ 1 + \kappa +
{\cal
O}(\kappa^2) \biggr],
\nonumber\\
&&\tau_{zz}(\vec{k}) = \frac{ B_0^2 }{\pi^2
L_{s}^4 k_s^3} \sqrt{\frac{\pi}{2}}\frac{5-3 n^2}{20} \biggl[ 1 +
\kappa +
{\cal O}(\kappa^2) \biggr],
\nonumber\\
&& \tau_{x z}(\vec{k}) = \frac{ B_0^2 }{\pi^2 L_{s}^4 k_s^3}
{\cal O}(\kappa),~~~
\tau_{yz}(\vec{k}) =  \frac{ B_0^2 }{\pi^2 L_{s}^4 k_s^3}{\cal O}
(\kappa),~~~
\tau_{x y} (\vec{k})=  \frac{ B_0^2 }{\pi^2 L_{s}^4 k_s^3}{\cal
O}(\kappa^2),
\label{limit}
\end{eqnarray}
where ${\cal O}$ is the Landau symbol.
We see that in the infrared limit ($\kappa <1$) the off-diagonal
terms
in the energy-momentum tensor are smaller than the diagonal ones. Of
course at very small scales (smaller than the size of the knot ) the
diagonal and off-diagonal terms are damped in the same way.
We will use often the following notation
\begin{eqnarray}
&&\tau_{00}(\vec{k}) =  \frac{ B_0^2 }{\pi^2
L_{s}^4 k_s^3} \sqrt{\frac{\pi}{2}}\frac{n^2 + 1}{4} \biggl[ 1 +
\kappa + {\cal
O}(\kappa^2) \biggr]e^{- \kappa},
\nonumber\\
&&\tau_{xx}(\vec{k}) \sim \tau_{yy}(\vec{k}) = \frac{ B_0^2 }{\pi^2
L_{s}^4 k^3_{s}} \sqrt{\frac{\pi}{2}}\frac{n^2}{5} \biggl[ 1 + \kappa
+ {\cal
O}(\kappa^2) \biggr]e^{- \kappa},
\nonumber\\
&& \tau_{zz}(\vec{k}) =  \frac{ B_0^2 }{\pi^2
L_{s}^4 k^3_{s}} \sqrt{\frac{\pi}{2}}\biggl(\frac{5 - 3
n^2}{20}\biggr) \biggl[ 1 + \kappa + {\cal O}(\kappa^2) \biggr]e^{- \kappa},
\label{inf}
\end{eqnarray}
where we kept (in square brackets) the leading infra-red contribution
and (outside the brackets) the leading (ultra-violet) damping term.
The mathematical reason for this notation is very simple. In doing
the
ultra-violet expansion of the Fourier transformed components of
$\tau_{ij}(\vec{k})$ we can clearly see that the leading exponential
damping factorises in the sense that each term appearing in the
various components of $\tau_{ij}$ is always multiplied by one of the
different functions reported in Tab. I. The common
feature
of all these functions is the exponential damping which becomes
effective at small scales and which is connected with the analytical
form of the components of $\vec{H}$ which (in real space) all share
the $1/( {\cal R}^2 + 1)^3$ behaviour.
Notice finally that, order by order in $\kappa$ the trace of
$\tau_{\mu\nu}$ vanishes. To lowest order, from Eqs. (\ref{inf}) one
can immediately check that $\tau_{\lambda}^{\lambda} = 0 + {\cal
O}(\kappa)$.
\begin{table}
\begin{tabular}{|l|l|}
\hline
$ I_{1}(\kappa)=\sqrt{\frac{\pi}{2}} \frac{1}{\kappa} e^{-
 \kappa}$&
$I_{2}(\kappa)=\frac{1}{2} \sqrt{\frac{\pi}{2}} e^{- \kappa}$\\
\hline
$I_{3}(\kappa)=\frac{1}{2}\sqrt{\frac{\pi}{2}}( 1 +\kappa) e^{-
 \kappa}$ &
$I_{4}(\kappa)=\frac{1}{48} \sqrt{\frac{\pi}{2}} ( 3 + 3 \kappa +
 \kappa^2) e^{- \kappa}$\\
\hline
$I_{5}(\kappa)=\frac{1}{384} \sqrt{\frac{\pi}{2}} ( 15 + 15 \kappa +
6
 \kappa^2 + \kappa^3) e^{- \kappa}$ &
$I_{6}(\kappa)=\frac{1}{3840}\sqrt{\frac{\pi}{2}}( 105 + 105 \kappa +
 45 \kappa^2 + 103 \kappa^3 + \kappa^4)e^{- \kappa}$\\
\hline
\end{tabular}
\label{Tab1}
\caption{We report the coefficients appearing in the Fourier
transform
of the Cartesian components of the energy-momentum tensor. We recall
that, in our notations, $\kappa = |\vec{k}|/k_{s}$.}
\end{table}
\end{appendix}

\end{document}